\documentclass[aps,twocolumn,showpacs,prl,floatfix,superscriptaddress,longbibliography]{revtex4-1}

\usepackage{color}
\usepackage{bm}
\usepackage{graphicx}
\usepackage{amsmath,amssymb}
\usepackage{newtxtext,newtxmath}
\usepackage{accents}
\usepackage{float}
\usepackage{wrapfig}
\usepackage{layout}
\usepackage{braket}
\let\%\relax 
\DeclareFontFamily{OT1}{pxr}{}
\DeclareFontShape{OT1}{pxr}{m}{n}{<->pxr}{}
\DeclareSymbolFont{letA}{OT1}{pxr}{m}{n}
\DeclareMathSymbol{\%}{0}{letA}{`\%}
\usepackage{scalerel,stackengine}
\stackMath
\newcommand\reallywidecheck[1]{%
\savestack{\tmpbox}{\stretchto{%
  \scaleto{%
    \scalerel*[\widthof{\ensuremath{#1}}]{\kern-.6pt\bigwedge\kern-.6pt}%
    {\rule[-\textheight/2]{1ex}{\textheight}}%WIDTH-LIMITED BIG WEDGE
  }{\textheight}% 
}{0.5ex}}%
\stackon[1pt]{#1}{\scalebox{-1}{\tmpbox}}%
}
\newcommand{\ie}{i.e.}
\newcommand{\tPt}{${}^{3}P_{2}$ }

\newcommand{\ii}{i}
\newcommand{\F}{\mathrm{F}}

\newcommand{\BdG}{\mathrm{BdG}}

\newcommand{\n}{\mathrm{n}}

\usepackage[normalem]{ulem}

\begin{document}
\preprint{aps/}
\title{Non-Abelian Half-Quantum Vortices in $^{3}P_{2}$ Topological Superfluids}

\author{Yusuke Masaki}
\email{MASAKI.Yusuke@nims.go.jp}
\affiliation{Research and Education Center for Natural Sciences, Keio University, Hiyoshi 4-1-1,
Yokohama, Kanagawa 223-8521, Japan}
\affiliation{Department of Physics, Tohoku University, Sendai, Miyagi 980-8578, Japan}
\affiliation{International Center for Materials Nanoarchitectonics, National Institute for Materials Science, Tsukuba, Ibaraki 305-0044, Japan}
\author{Takeshi Mizushima}
\affiliation{Department of Materials Engineering Science, Osaka University, Toyonaka, Osaka 560-8531, Japan}
\author{Muneto Nitta}
\affiliation{Research and Education Center for Natural Sciences, Keio University, Hiyoshi 4-1-1,
Yokohama, Kanagawa 223-8521, Japan}
\affiliation{Department of Physics, Keio University, Hiyoshi 4-1-1, Japan}
\date{\today}
\begin{abstract}
\tPt superfluids realized in neutron stars
are the largest topological quantum matters in our Universe.  We establish the existence and stability of non-Abelian half-quantum vortices (HQVs) in \tPt superfluids with strong magnetic fields. Using a self-consistent microscopic approach, we find that a singly quantized vortex is energetically destabilized into a pair of two non-Abelian HQVs owing to the strongly spin-orbit-coupled pairing. We find a topologically protected Majorana fermion on each HQV, thereby providing two-fold non-Abelian anyons characterized by both Majorana fermions and a non-Abelian first homotopy group.
\end{abstract}

\maketitle

{\it Introduction.}---
Quantum physics tells us that 
all particles are either fermions or bosons 
under certain assumptions; 
a wave function of multi-particle states is  
symmetric (asymmetric) under 
the exchange of two bosons (fermions).
However, an exception, anyons, exists.
The exchange of two anyons causes the wave function to acquire a phase factor~\cite{Leinaas:1977fm,Wilczek:1982wy}.
Such anyons explain the physics of fractional quantum Hall states~\cite{Halperin:1984fn,Arovas:1984qr}, 
and have been experimentally observed for 
a $\nu=1/3$ fractional quantum Hall state~\cite{Nakamura:2020}.
Recently, another option has attracted great attention, that is, {\it non-Abelian anyons}.
The exchange of two non-Abelian anyons  
leads to a unitary matrix acting on a set of wave functions 
as a generalization of the phase factor for 
Abelian anyons. 
Although non-Abelian anyons have yet to be observed, 
they have been theoretically predicted to exist in 
 $\nu=5/2$ fractional quantum Hall states~\cite{Moore:1991ks}, 
 topological superconductors~\cite{Read:1999fn,Ivanov:2000mjr},
 and spin liquids~\cite{Kitaev2006,Motome2020}.
Non-Abelian anyons have attracted significant interest owing to the possibility for a platform of topological quantum computation~\cite{Kitaev:1997wr,Nayak:2008zza,sarma_majorana_2015} 
which are robust against noise, 
in contrast to the conventional quantum computation methods.

There are two apparentlydifferent origins of non-Abelian anyons, 
one fermionic and the other bosonic.
The fermionic origin is based on Majorana fermions
realized in topological superconductors~\cite{Ivanov:2000mjr,Kitaev:2001kla,alicea2012,Leijnse2012,Beenakker2013,Elliott2015,Sato2016,Beenakker2020}.
Majorana fermions are particles that coincide with 
their own anti-particles~\cite{majorana_teoria_1937}. 
This is the main route 
for topological quantum computation. 
By contrast, 
non-Abelian anyons are also realized in bosonic systems, the statistics of which are due to non-Abelian vortices supported by 
a non-Abelian first homotopy group
of order parameter (OP) manifolds, 
giving non-commutativity under the exchange of two vortices~\cite{Lo:1993hp,PhysRevLett.123.140404}. 
Examples can be found in 
liquid crystals~\cite{Poenaru1977,Mermin:1979zz} 
and spinor Bose-Einstein condensates (BECs)~\cite{Semenoff:2006vv,Kobayashi:2008pk,Borgh:2016cco,PhysRevLett.123.140404}.
Two apparently different non-Abelian anyons 
have been discussed separately thus far 
and their relation has yet to be clarified.

\begin{figure}[b!]
\includegraphics[width=\hsize]{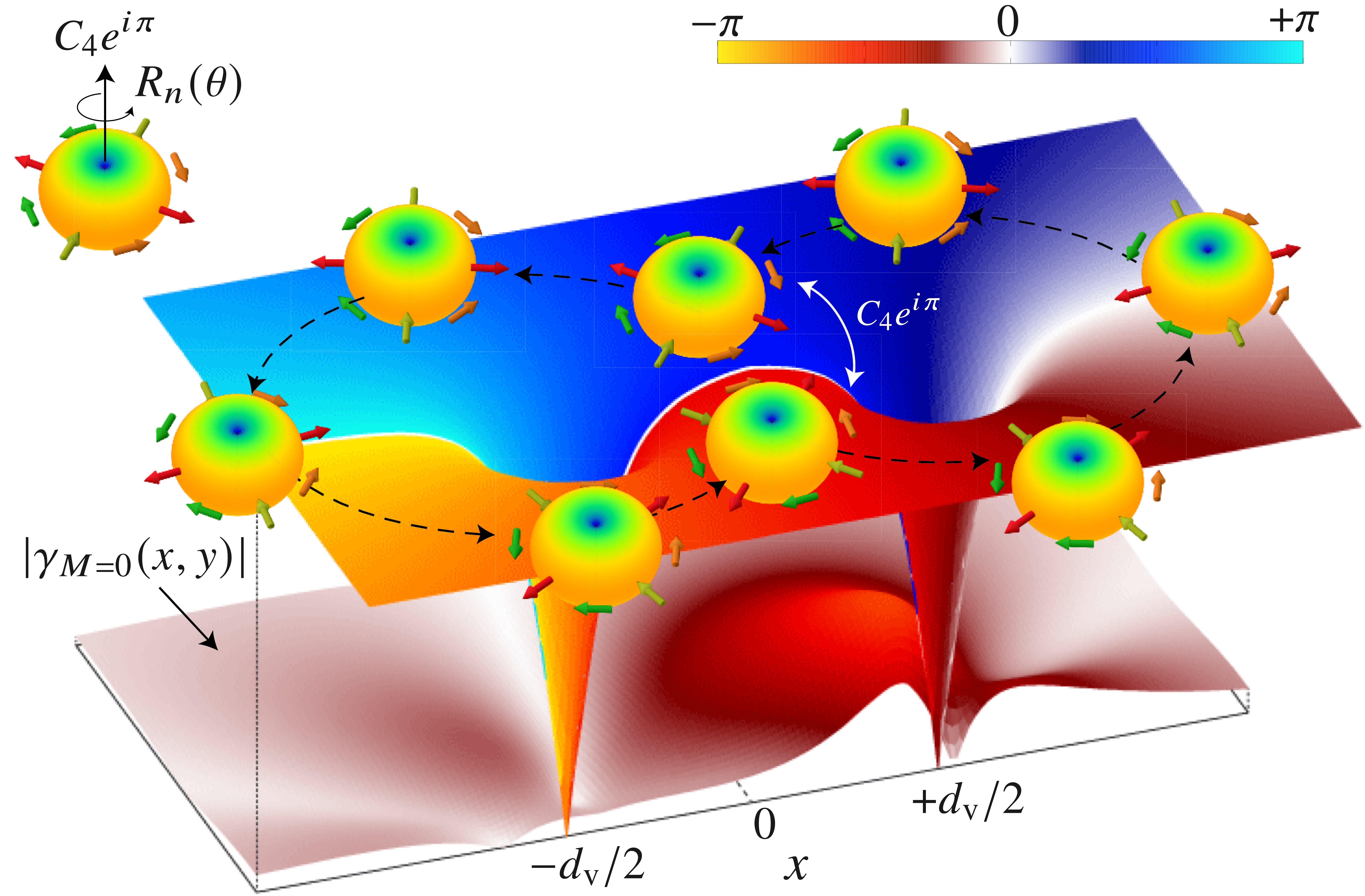}
\caption{Schematic of a pair of non-Abelian HQVs $(\kappa,n)=(1/2,+1/4)$ at $x=d_{\mathrm{v}}/2$ and $(1/2,-1/4)$ at $x=-d_{\mathrm{v}}/2$ in a $D_{4}$-BN state. Its spin-momentum structure is shown by objects with color arrows representing $d$ vectors.}
\label{fig:hqv_image}
\end{figure}

The aim of this Letter is to present vortices 
simultaneously accompanied by 
the two different non-Abelian natures, that is, fermionic and bosonic origins of 
non-Abelian anyons. 
A system that realizes such vortices
is a neutron superfluid 
expected to occur in neutron star cores.
This is called the \tPt superfluid 
(spin-triplet $p$-wave pairing) of neutrons~\cite{Hoffberg:1970vqj,Tamagaki1970,Takatsuka1971,Takatsuka1972,Richardson:1972xn,Sedrakian:2018ydt}, 
which has been recently shown to be 
the largest topological quantum matter 
in our Universe~\cite{Mizushima:2016fbn}  
(a class DIII in the classification of topological insulators and superconductors~\cite{Schnyder:2008tya,Ryu:2010zza}), allowing a gapless Majorana fermion on its boundary~\cite{Mizushima:2016fbn} 
and vortex cores~\cite{Masaki:2019rsz}. 
From the Ginzburg--Landau (GL) theory~\cite{Richardson:1972xn,Fujita1972,Sauls:1978lna,Muzikar:1980as,Sauls:1982ie, Yasui:2018tcr,Yasui:2019unp},  
this matter was found to admit 
non-Abelian half-quantum vortices (HQVs)~\cite{Sauls1980,Masuda:2016vak}
in addition to integer vortices~\cite{Richardson:1972xn,Fujita1972,Sauls:1978lna,Muzikar:1980as,
Sauls:1982ie,Masuda:2015jka,Chatterjee:2016gpm},  
coreless vortices~\cite{Leinson:2020xjz},
domain walls \cite{Yasui:2019vci}, 
and boojums on the surface
\cite{Yasui:2019pgb}.
Such topological defects may play a crucial role 
in the dynamics and evolution of neutron stars. 
In particular, 
the existence of HQVs was proposed to explain 
a longstanding unsolved problem of neutron stars: 
the origin of the pulsar glitch phenomena, that is, sudden speed-up events of neutron stars~\cite{Marmorini:2020zfp}.
Unlike the Feynman--Onsager's quantization of circulation, 
HQVs or more generally, fractionally quantized vortices~\cite{Babaev:2001hv,Babaev:2004rm,Babaev:2007}, appear ubiquitously in diverse systems 
with multiple components.  
The topological stability of HQVs (or fractional quantum vortices) has been predicted in the A phase~\cite{PhysRevLett.55.1184,salomaaRMP} 
of superfluid $^3$He, 
unconventional superconductors~\cite{Kee2000,Chung:2007zzc,Garaud2012a,Zyuzin2017,Etter2020,How2020}, 
spinor BECs~\cite{
Leonhardt:2000km,
Ho:1998zz,doi:10.1143/JPSJ.67.1822,Shinn:2018zde,
Semenoff:2006vv,Kobayashi:2008pk,2010arXiv1001.2072K,Borgh:2016cco}, multicomponent superconductors~\cite{Goryo2007,Tanaka2007,Crisan2007,Guikema2008,
Tanaka2017,Tanaka2018} and 
BECs~\cite{Son2002,Kasamatsu2004,Eto2011,Eto:2012rc,
Cipriani2013,Tylutki2016,Kasamatsu2016,
MenciaUranga2018,Eto2018,Kobayashi:2018ezm},  
and even high-energy physics such as 
quantum chromodynamics~\cite{Balachandran:2005ev,Nakano:2007dr,
Eto:2013hoa,Eto:2021nle,Fujimoto:2020dsa} 
and 
physics beyond the Standard Model of elementary particles~\cite{Dvali:1993sg,Eto:2018tnk}.
{\it Abelian} HQVs were experimentally confirmed  
in the uniaxially disordered superfluid $^{3}$He~\cite{Autti:2015bta,Makinen:2018ltj} 
and in a spinor BEC~\cite{PhysRevLett.115.015301}.
However, no systems admitting {\it non-Abelian} HQVs 
with Majorana fermions have been known thus far. 

In this study, we microscopically 
establish the existence and stability of 
non-Abelian HQVs, along each of which, we find a topologically protected gapless
Majorana fermion.
In the presence of a strong magnetic field relevant for magnetars, 
i.e., neutron stars accompanied by extraordinary 
large magnetic fields,  
the ground state is in a dihedral-four biaxial nematic ($D_{4}$-BN) phase~\cite{Masuda:2015jka,Mizushima:2016fbn,Mizushima:2019spl}. 
There, a singly vortex 
is shown to be split into two non-Abelian HQVs.
Each HQV admits a gapless Majorana fermion, 
thereby being a new type of non-Abelian anyons.
We also calculate the interaction energy between HQVs
and find an intrinsic mechanism of their thermodynamic stability due to the uniaxial nematic pairing induced around the cores.

%%%%%%%%%%%%%%%%%%%%%%%%

{\it Non-Abelian HQVs.}---
Here, we focus on non-Abelian HQVs in the $D_{4}$-BN phase of a \tPt superfluid. Let us consider systems invariant under a $\mathrm{U}(1)$ gauge transformation and $\mathrm{SO}(3)$ spin-momentum rotation. 
A \tPt superfluid is the condensation of spin-triplet Cooper pairs with a total angular momentum of $J=2$, the OP of which is given by a $3\times 3$ traceless symmetric tensor, $\mathcal{A}_{\mu \nu (\mu,\nu=x,y,z)}$, with spin index $\mu$ and momentum index $\nu$. 
The continuous symmetries act as $\mathcal{A} \rightarrow e^{i\varphi}g\mathcal{A}g^{\mathrm{tr}}$, $e^{i\varphi}\in \mathrm{U}(1)$ and $g\in\mathrm{SO}(3)$. 
The homogeneous OP of the $D_{4}$-BN state has a diagonal form~\cite{Sauls:1978lna}:
\begin{equation}
\mathcal{A}_{\mu \nu} = \Delta\begin{pmatrix}
1 & 0 & 0 \\ 0 & -1 & 0 \\ 0 & 0 & 0
\end{pmatrix}_{\mu \nu},
\end{equation}
which is invariant under a $C_{4}$ rotation around the $z$ axis in a point node direction, combined with the $\pi$ phase rotation. 
Its spin-momentum structure is schematically shown by $d$ vectors, $d_\mu(\bm{k}) = \sum_{\nu}\mathcal{A}_{\mu\nu} k_{\nu}$, using arrows in the top-left object in Fig.~\ref{fig:hqv_image}. 
A large magnetic field relevant to magnetars thermodynamically stabilizes the $D_{4}$-BN state with point nodes along the direction of the magnetic field~\cite{Masuda:2015jka,Mizushima:2016fbn,Yasui:2018tcr,Mizushima:2019spl}. 

The OP manifold in the $D_{4}$-BN state, $R=[\mathrm{U}(1)\times \mathrm{SO}(3)]/D_{4}$, leads to rich topological charges of line defects supported by the first homotopy group 
$\pi_{1}(R) = \mathbb{Z}\times_{h} D_{4}^{\ast}$~\cite{Kobayashi:2011xb}. 
This includes non-Abelian HQVs~\cite{Masuda:2016vak}, which are vortices with noncommutative topological charges.
An asymptotic form for an isolated vortex is given by
\begin{equation}
\mathcal{A}(\theta) =  e^{i\kappa\theta}R_{n}(\theta) \mathcal{A}
R^{\mathrm{tr}}_{n}(\theta),
\label{eq:bc}
\end{equation}
where $\theta\equiv \tan^{-1}(y/x)$ is the azimuthal angle and $R_{n}(\theta)\in \mathrm{SO}(2)$ is a rotation matrix around the $z$ axis by the angle $n\theta$. The integer vortices are characterized by $\kappa \in \mathbb{Z}$ and $n = 0$. In the $D_{4}$-BN state, the $\pi$ phase jump arising from $\kappa\in \mathbb{Z}+1/2$ is compensated by the $C_{4}$ rotation with $n \in \mathbb{Z} \pm 1/4$. 
Thus, HQVs are topologically allowed, and a singly quantized vortex is predicted to be split into  a pair of HQVs, as  illustrated in Fig.~\ref{fig:hqv_image}, 
where HQVs with $(\kappa,n)=(1/2,-1/4)$ and 
$(1/2,+1/4)$ are placed 
at $x=-d_{\mathrm{v}}/2$ and
 $x=d_{\mathrm{v}}/2$, respectively.

{\it Structure and Stability of the HQVs.}---
To microscopically discuss the stability of non-Abelian HQVs, we utilize quasiclassical theory. A fundamental quantity is the quasiclassical propagator, $\check{g}(\bm{k}_{\F},\bm{R};\ii\omega_{n})$, with the Fermi momentum $\bm{k}_{\F}$, governed by the Eilenberger equation~\cite{eilenberger,serene,SM},
\begin{equation}
0 = \ii \bm{v}_{\F}\cdot \bm{\nabla} \check{g}+ [\ii\omega_{m}\check{\tau}_{z} + \check{u} +\check{\sigma}_{\Delta}, \check{g}],
\label{eq:E}
\end{equation}
where the symbol $\check{\cdot}$ denotes a $4\times4$ matrix in the spin and Nambu space, $\bm{v}_{\F}$ is the Fermi velocity, and $\omega_{m} = \pi T (2m + 1)$ is the fermionic Matsubara frequency ($m\in \mathbb{Z}$). 
The Zeeman field $V_{\mathrm{Z}}$ along the $z$ axis is introduced through $\check{u} = V_{Z}\mathrm{diag}(\hat{\sigma}_{z},\hat{\sigma}_{z}^{\mathrm{tr}})$. 
The Pauli matrices in the Nambu space and the spin space are also introduced as $\check{\tau}_{\mu}$ and $\hat{\sigma}_{\mu}$, respectively, where $\hat{\cdot}$ denotes a $2\times2$ matrix in the spin space.
The self-energy matrix $\check{\sigma}_{\Delta}$ is composed of the \tPt OP $\mathcal{A}_{\mu\nu}$. Assuming uniformity along the $z$-direction, we determine the spatial profile of $\mathcal{A}_{\mu\nu}(\bm{R}=(x,y))$ 
by self-consistently solving Eq.~\eqref{eq:E} complemented with a gap equation for interacting neutrons through a zero-range attractive \tPt force (see Ref.~\cite{SM} for the detail). Below, we show the numerical results at $T = 0.4 T_{\mathrm{c}}$ and $V_{\mathrm{Z}} = 0.5 T_{\mathrm{c}}$ with the critical temperature $T_{\mathrm{c}}$. For this parameter set, the $D_{4}$-BN state is the most stable uniform state.

\begin{figure}[t]
\begin{center}
\includegraphics[width=\hsize]{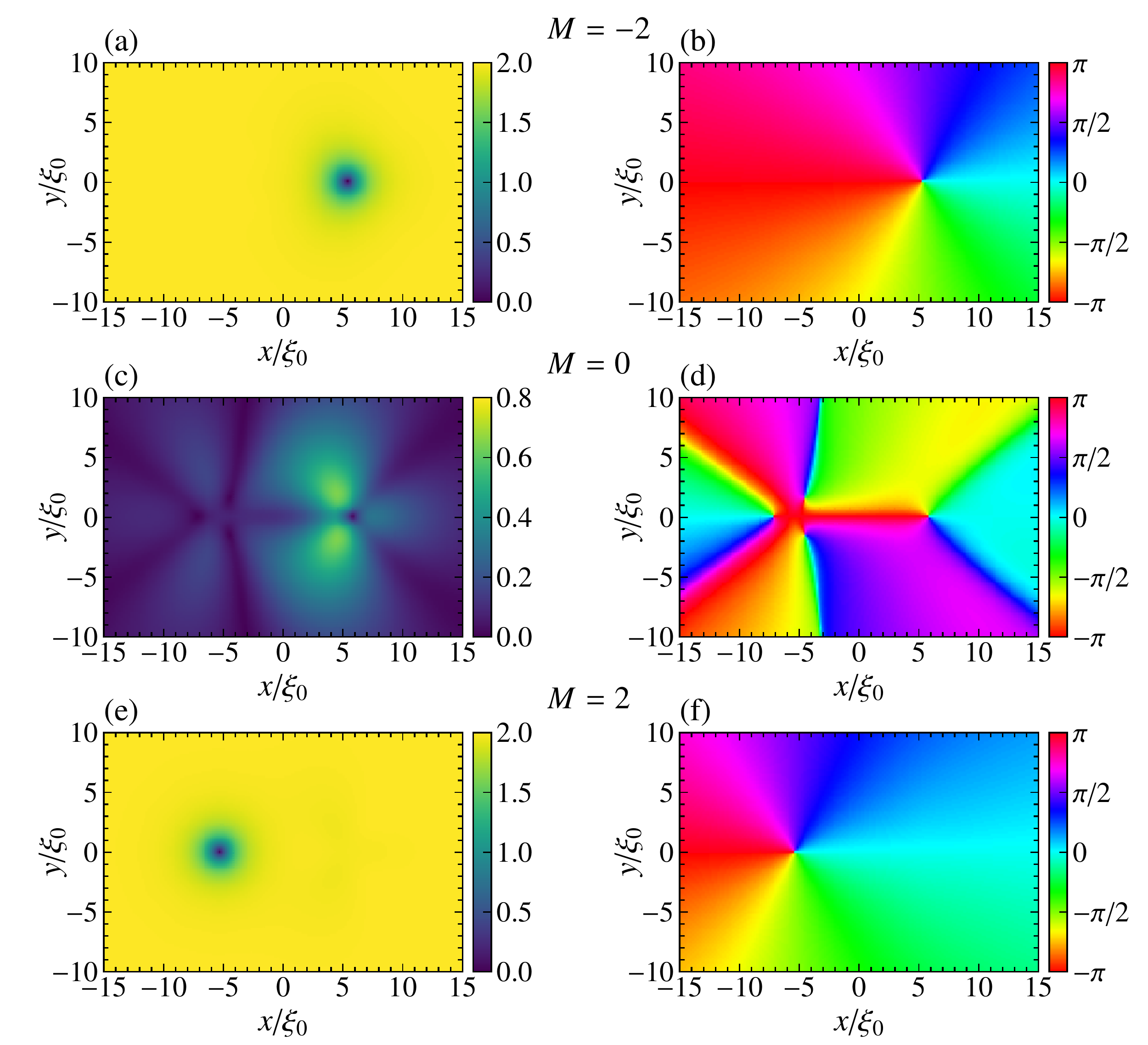}
\caption{Two HQVs with $d_{\mathrm{v}} \simeq 10.7\xi_{0}$. Panels (a), (c), and (e) show the spatial profiles of the amplitude $|\gamma_{M=-2,0,2}(\bm{R})|$, whereas panels (b), (d), and (f) show those of the phase $\mathrm{arg}[\gamma_{M=-2,0,2}(\bm{R})]$. The middle panels (c) and (d) represent the induced component.}
\label{fig:hqvs}
\end{center}
\end{figure}

In Fig.~\ref{fig:hqvs}, we show a pair of HQVs with finite intervortex distance $d_{\mathrm{v}} \simeq 10.7 \xi_{0}$, where $\xi_{0}=\varv_{\F}/(2\pi T_{\mathrm{c}})$ is the coherence length. 
It is convenient to expand $\mathcal{A}_{\mu\nu}$ as $\mathcal{A}_{\mu\nu} = \sum_{M=-2}^{2} \gamma_{M}(\bm{R})\Gamma_{M,\mu\nu}$,
where $\Gamma_{M}$ is a $3\times3$ basis tensor of the $z$ component of the total angular momentum $J_{z}$ such that $J_{z} \Gamma_{M} = M \Gamma_{M}$
and $\gamma_{M}(\bm{R})$ is the complex OP projected onto the sector $J_{z} = M$. The $D_{4}$-BN state is represented by $|\gamma_{M=2}| = |\gamma_{M=-2}|$ and $\gamma_{M=-1,0,1} = 0$. For an isolated HQV, the asymptotic form of Eq.~\eqref{eq:bc} is recast into $\mathcal{A}_{\mu\nu}(\theta) = \sum_{M=-2,2} \gamma_{M}e^{\ii ( \kappa \theta - M\varphi)} \Gamma_{M,\mu\nu}$ with the vorticity $\kappa = 1/2$ and the rotation angle of the triad $\varphi = n\theta = \pm \theta/4$. We set $\kappa>0$ without a loss of generality, whereas the choice of $n=+1/4$ ($-1/4$) corresponds to the clockwise (counterclockwise) texture of the gap structure. The two HQVs shown in Figs.~\ref{fig:hqv_image} and \ref{fig:hqvs} are characterized by a pair of $(\kappa,n)=(1/2,-1/4)$ at $x=-d_{\mathrm{v}}/2$ and $(1/2,+1/4)$ at $x=d_{\mathrm{v}}/2$. 
The amplitudes (phases) of $\gamma_M(\bm{R})$ are shown in the left (right) panels of Fig.~\ref{fig:hqvs}. 
In each $M=\pm2$ sector, a single winding structure is realized [panels (b) and (f)], and in the $M=0$ sector, a structure with a winding of $2 = 3 - 1$ is induced, as indicated in panel (d). Note that in the bulk region, $\gamma_{M=0}$ moves toward zero.

%%%%%%%%%%%%%%%%%%%%%%%%

The isolated HQV for $n = + 1/4$~$(-1/4)$ consists of three components, that is, a singular vortex component for $M = -2$~($+2$), an almost uniform unwinding component for $M=+2$~($-2$), and the induced component for $M=0$.
It can be regarded as a chiral $p$-wave superconducting vortex with the spin parallel to the chirality, and the phase windings of the induced components are $-1$ for $(\kappa,n)= (1/2, 1/4)$ and $3$ for $(1/2,-1/4)$~\cite{Heeb1999,MatsumotoHeeb2001}. In the former (latter) case, the vorticity is antiparallel (parallel) to the chirality. However, the amplitude of the induced component $|\gamma_{M=0}(\bm{R})|$ breaks the axial symmetry to a 3-fold symmetry for $n = 1/4$ and 5-fold symmetry for $n = -1/4$ (see Ref.~\cite{SM}). The axial symmetry is also broken by the boundary conditions. 

The two types of internal structures in the $M=0$ component induced for HQVs with $n = \pm1/4$ are modulated by the connection of these two HQVs. We find that this modulation causes an interaction between the two HQVs and binds them together.
To unveil the interaction between HQVs, we compute the Luttinger--Ward energy functional $\mathcal{J}_{\mathrm{sn}}$ from the self-consistently determined $\check{g}$~\cite{Vorontsov2003}.
For several values of $d_{\mathrm{v}}$, we calculate the interaction energy in the following steps. 
We construct three solutions: one includes two HQVs with their centers at $\bm{R}_{1}=(d_\mathrm{v}/2,0)$ and $\bm{R}_{2}=(-d_\mathrm{v}/2,0)$, as shown in Fig.~\ref{fig:hqvs}, and this energy is denoted by $\mathcal{J}_{\mathrm{sn}}(\bm{R}_{1},\bm{R}_{2})$. The other two are the corresponding isolated HQVs $(\kappa,n)=(1/2, 1/4)$ and $(1/2,-1/4)$, whose centers are at $\bm{R}_{1}$ and  $\bm{R}_{2}$, respectively. The energies of these two solutions are $\mathcal{J}_{\mathrm{sn}}^{+}(\bm{R}_{1})$ and  $\mathcal{J}_{\mathrm{sn}}^{-}(\bm{R}_{2})$. 
The interaction energy is defined by $
\Delta \mathcal{J}_{\mathrm{sn}}(d_{\mathrm{v}}) = 
\mathcal{J}_{\mathrm{sn}}(\bm{R}_{1},\bm{R}_{2}) - 
\mathcal{J}_{\mathrm{sn}}^{+}(\bm{R}_{1}) - 
\mathcal{J}_{\mathrm{sn}}^{-}(\bm{R}_{2})$. 
We will now remark on the boundary effects. The long-tailed flows of the mass and spin currents are cut off owing to a finite-sized simulation box, so that the translational symmetry for a single HQV is broken. Similar effects are also included in the two-HQV solution and its energy. 
In our strategy, 
the boundary contributions in $\mathcal{J}_{\mathrm{sn}}$
cancel out those in $\mathcal{J}_{\mathrm{sn}}^{\pm}$
and only the interaction energy of the two HQVs becomes available. 

\begin{figure}[t]
\begin{center}
\includegraphics[width=\hsize]{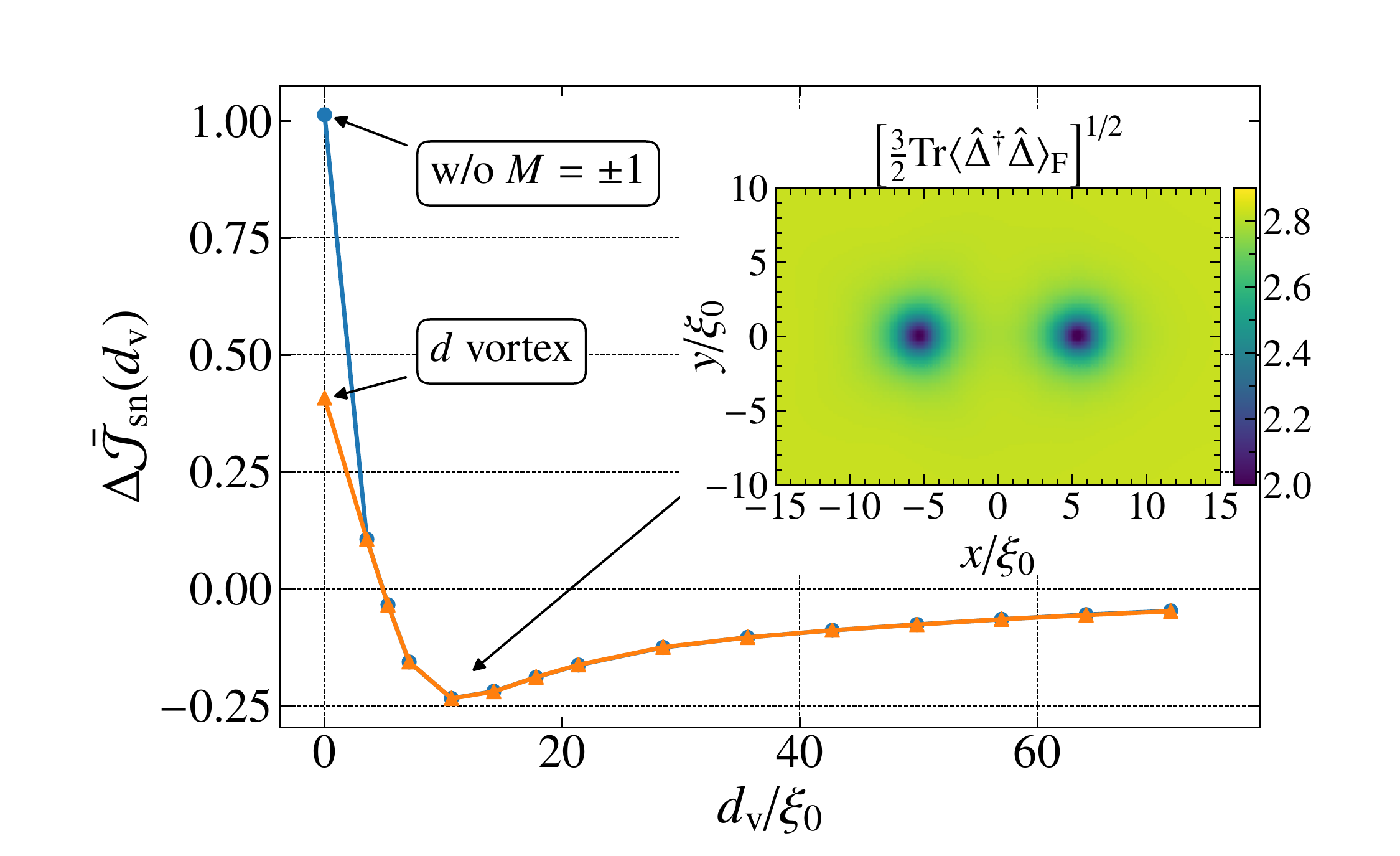}
\caption{Interaction energy of two HQVs as a function of their separation $d_{\mathrm{v}}$. 
The triangular (circular) symbols are calculated by considering (neglecting) the possibilities of the induced components of $M=\pm1$.
The inset shows the total amplitude of the OP 
for the OP shown in Fig.~\ref{fig:hqvs}, and the intervortex distance is indicated by the arrow. The free energy is scaled as $\bar{\mathcal{J}}_{\mathrm{sn}} = \mathcal{J}_{\mathrm{sn}}/(\nu_{\n} T_{\mathrm{c}}^{2} \xi_{0}^{2} \Omega_{z})$, where $\Omega_{z}$ is the length of the system in the $z$-direction, and $\nu_{\n}$ is the density of states at the Fermi energy in the normal state.
}
\label{fig:free-energy}
\end{center}
\end{figure}
In Fig.~\ref{fig:free-energy}, we show the interaction energies of the two HQVs calculated. The triangular (circular) symbols are calculated by considering (neglecting) the induced components $\gamma_{M = \pm 1}$. The difference appears only for $d_{\mathrm{v}} = 0$, where single integer vortices are realized; The triangular symbol at $d_{\mathrm{v}} = 0$ stands for the double-core vortex ($d$ vortex), whose core is occupied with $\gamma_{M=\pm 1}$, as in the superfluid ${}^{3}$He B phase~\cite{thunebergPRL86,Salomaa1986}. The $d$ vortex (triangular symbol) has the lower energy than the vortex without $\gamma_{M=\pm1}$ (circular symbol) 
because condensation energy due to $\gamma_{M=\pm1}$ is gained at the origin.
Significantly, for finite $d_{\mathrm{v}}$, the interaction energy decreases as $d_{\mathrm{v}}$ increases from zero and reaches the minimum at a finite intervortex distance $d_{\mathrm{v}}$, which means that the $d$ vortex is unstable for splitting into the two HQVs. The gain in the interaction energy is due to the deformations in $\gamma_{M=0}(\bm{R})$, and the two HQVs form a bound molecule with an optimal separation. 

Molecules of HQVs are also discussed in
 superfluid ${}^{3}$He~\cite{PhysRevLett.55.1184} and unconventional superconductors~\cite{Chung:2007zzc},
but their stabilization mechanisms are different from ours: 
In the superfluid ${}^{3}$He-A phase, the spin mass correction through the Fermi liquid correction was phenomenologically introduced to stabilize the HQV~\cite{PhysRevLett.55.1184}; however, its realization remains controversial because the strong coupling effects destabilize the HQV~\cite{Kawakami2009,Kawakami2010,Kawakami2011,mizushimaJPSJ16}. In the polar phases, the stability of the HQVs is  supported  by
an extrinsic mechanism from strong anisotropic impurity effects  using the GL theory~\cite{Nagamura2018,Tange2020, Regan2021arxiv}. 
There is no intrinsic interaction between the two HQVs in the weak coupling limit because two spin sectors are independent. By contrast, in the present case, a new mechanism of the interaction originates from the deformation in the induced component $\gamma_{0}$ because of the strongly spin-orbit-coupled pairing.

\begin{figure}[t]
\begin{center}
\includegraphics[width=\hsize]{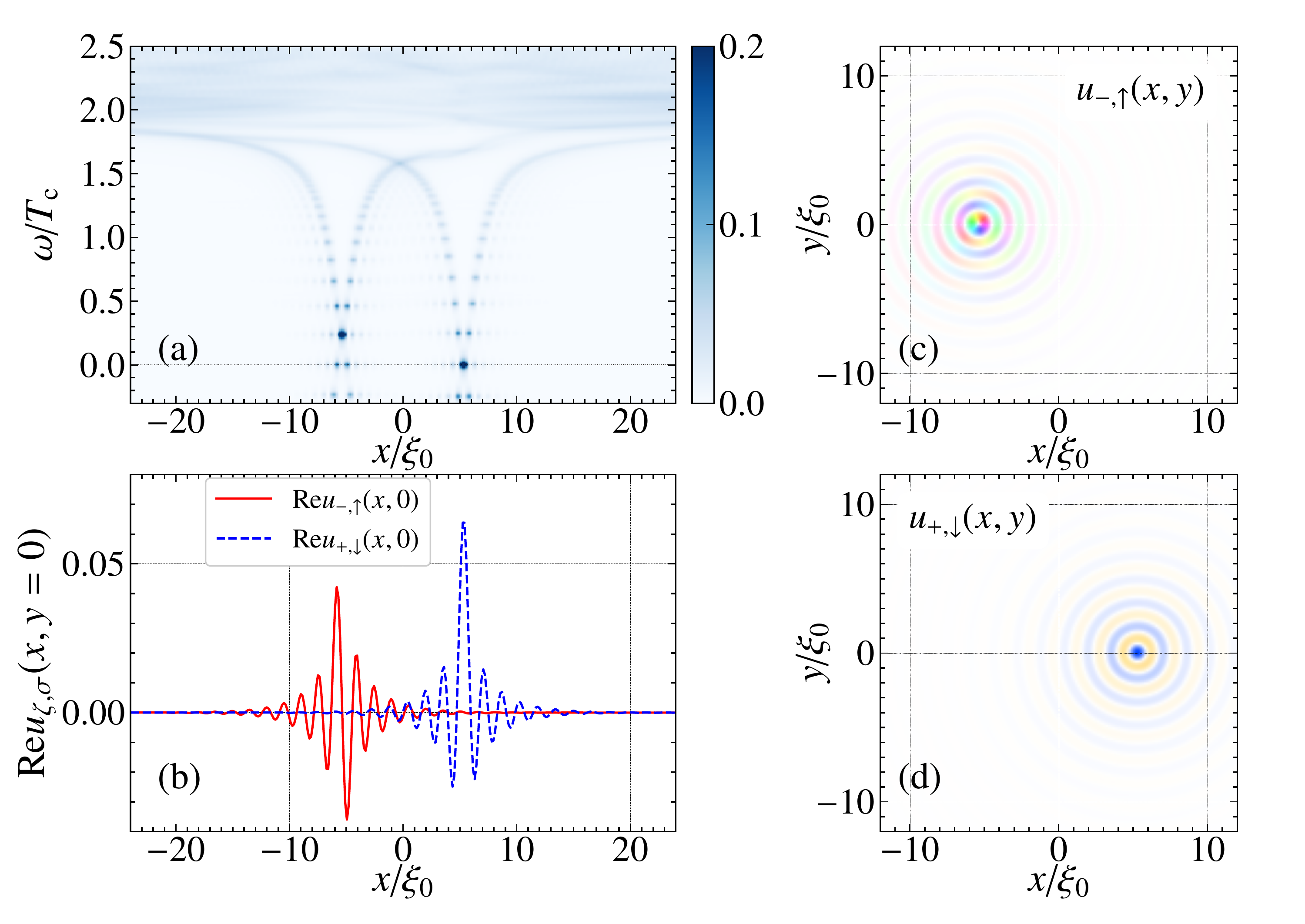}
\caption{
	(a) Local density of states $\nu_{k_{z} = 0}(\bm{R};\omega)$ at $k_{z} = 0$ and $y = 0$ for a pair of HQVs located at $(x,y)=(\pm d_{\mathrm{v}}/2,0)$ with $d_{\mathrm{v}} \simeq10.7\xi_0$.
	(b) The real parts of two Majorana wave functions $u_{-,\uparrow}$ and $u_{+,\downarrow}$ are shown along the $x$ axis.
	(c) (d) Two-dimensional spatial profile of $u_{-,\uparrow}$ (c) and $u_{+,\downarrow}$ (d). The color maps indicate their phase information, and the color bar is the same as in Fig.~\ref{fig:hqvs}(b). The intensities are indicated by the saturation.
}
\label{fig:ldos}
\end{center}
\end{figure}

{\it Majorana zero modes in non-Abelian HQVs.}---
Finally, we clarify the existence of topologically protected zero-energy states in HQVs, which behave as non-Abelian (Ising) anyons. Using the OP determined self-consistently for a separation of $d_{\mathrm{v}}\simeq 10.7 \xi_{0}$ and spatial uniformity along the $z$-direction, we solve the Bogoliubov--de Gennes (BdG) equation, $\check{\mathcal{H}}_{\BdG,k_{z}}(\bm{R}) \vec{u}_{\alpha,k_{z}}(\bm{R}) = \epsilon_{\alpha,k_{z}} \vec{u}_{\alpha,k_{z}}(\bm{R})$, 
where $\check{\mathcal{H}}_{\BdG,k_{z}}$ is a $4\times4$ matrix in the spin and Nambu space with the OP $\mathcal{A}_{\mu\nu}(\bm{R})$; $\vec{u}_{\alpha,k_{z}}(\bm{R}) = [u_{\alpha,k_{z},\uparrow}(\bm{R}),u_{\alpha,k_{z},\downarrow}(\bm{R}),\varv_{\alpha,k_{z},\uparrow}(\bm{R}),\varv_{\alpha,k_{z},\downarrow}(\bm{R})]$ is the $\alpha$-th eigenvector of the axial momentum $k_{z}$, and $\epsilon_{\alpha,k_{z}}$ is its eigenenergy. We set $k_{\F}\xi_{0} = 5$. 
For the spectroscopy of the vortex-bound states, we show the fermionic local density of states for $k_{z} = 0$ in Fig.~\ref{fig:ldos}(a), $\nu_{k_{z} =0}(\bm{R};\omega) = \sum_{\alpha,\sigma}|u_{\alpha,k_{z}=0,\sigma}(\bm{R})|^{2}\delta(\omega- \epsilon_{\alpha,k_{z}=0})$ along $y = 0$. 
In the energy region below the bulk gap $\omega_{\mathrm{g}}/T_{\mathrm{c}}\sim 2.0$, the spectral weights are localized around the HQV cores and the edge (not shown). 
The energy levels of the vortex bound states are discretized with level spacing on the order of $\omega^{2}_{\mathrm{g}}/\varepsilon_{\mathrm{F}}\sim 0.255 T_{\mathrm{c}}$. 

It is worth noting that each vortex hosts a single zero-energy state with numerical accuracy. Let us now clarify the symmetry protection and non-Abelian nature of the zero modes. For this purpose, we employ the semiclassical approximation as $\check{\mathcal{H}}_{\BdG,k_{z}}(x,y) \mapsto \check{\mathcal{H}}(\bm{k},\theta)$, which varies slowly in real-space coordinates. The spatial modulation due to a vortex line is considered as an adiabatic change in the Hamiltonian as a function of the azimuthal angle $\theta$ around the vortex line. 
For the topological protection of zero-energy states in a vortex, the mirror reflection $\mathcal{M}_{xy}$ with respect to the $xy$-plane is essential. As demonstrated in Refs.~\cite{ueno,tsutsumiJPSJ13}, if the gap function is odd under the mirror reflection, the HQV may support a Majorana zero mode protected by the mirror symmetry. 
For the mirror reflection invariant momentum $\bm{k}_{\mathrm{M}}\equiv(k_x,k_y,k_z=0)$, the BdG Hamiltonian commutes with $\check{\mathcal{M}}_{xy}^{-}=\mathrm{diag}(i\hat{\sigma}_z, i \hat{\sigma}_{z}^{\mathrm{tr}})$ as $[\check{\mathcal{H}}(\bm{k}_{\mathrm{M}},\theta),\check{\mathcal{M}}_{xy}^{-}]=0$, because $\mathcal{A}_{xz} = \mathcal{A}_{yz}=0$, \ie, $\gamma_{M=\pm 1} = 0$ in non-Abelian HQVs. Hence, the Hamiltonian with $k_{z}=0$ is block-diagonalized in terms of the eigenvalues of the mirror operator $\lambda = \pm i$, 
as 
$\check{\mathcal{H}}(\bm{k}_{\mathrm{M}},\theta) = \bigoplus_{\lambda}
\tilde{\mathcal{H}}_{\lambda}(\bm{k}_{\mathrm{M}},\theta)$, 
where the $2\times2$ submatrix $\tilde{\mathcal{H}}_{\lambda}$ is still subject to the particle-hole symmetry. 
In terms of the Altland-Zirnbauer symmetry classes, each subsector belongs to class D, similar to spinless chiral superconductors~\cite{Schnyder:2008tya}. 
The topological invariant relevant to the class-D BdG Hamiltonian, $\tilde{\mathcal{H}}_{\lambda}$, on the base space $(\bm{k}_{\mathrm{M}},\theta) \in S^{2}\times S^{1}$ is the $\mathbb{Z}_{2}$ number defined as~\cite{teo,qiPRB08}
\begin{equation}
\nu_{\lambda} = \left(\frac{i}{\pi} \right)^{2} \int _{S^{2}\times S^{1}} {\mathrm{tr}}[{A}d{A} + \frac{2}{3}{A}^{3}]~ \mod 2,
\end{equation}
with the Berry connection $A$ obtained from the occupied eigenstates of $\tilde{\mathcal{H}}_{\lambda}(\bm{k}_{\mathrm{M}},\theta)$ on the mirror-invariant plane.
The non-Abelian HQV in the $D_{4}$-BN state has a nontrivial value of the $\mathbb{Z}_{2}$ invariant in each mirror subsector,
$\nu_{\lambda} = +1$ ($-1$) for $\lambda = +i$ ($-i$).
The nontrivial (odd) values ensure 
a single Majorana zero mode in each HQV that behaves as a non-Abelian (Ising) anyon~\cite{sato14,tsutsumiJPSJ13}. In addition to the $\mathbb{Z}_2$ number, 
such a zero mode is protected by the winding number associated with the magnetic $\pi$ rotation~\cite{SM}.

In Figs.~\ref{fig:ldos}(b)--\ref{fig:ldos}(d),  we show the wave functions of the zero modes obtained by separating the edge mode and the vortex core mode through a linear combination of the two particle-hole symmetric eigenpartners. We assign the label $\zeta = +$ $(-)$ to the state localized around
$\bm{R}_{1}$ $(\bm{R}_{2})$ instead of $(\alpha, k_{z}= 0)$. By choosing the global phase of the wave function properly, the Majorana condition $u_{\zeta,\sigma}(\bm{R}) = (\varv_{\zeta,\sigma}(\bm{R}))^{*}$ is satisfied. The real parts of $u_{-,\uparrow}$ and $u_{+,\downarrow}$ along $y=0$ are shown in panel (b); for the other combinations of $\zeta$ and $\sigma$, the wave functions $u_{\zeta,\sigma}$ are zero. The phase windings of $u_{-,\uparrow}$ and $u_{+,\downarrow}$ are one and zero, as indicated by the two dimensional color maps in panels (c) and (d), respectively, for the same color bar in Fig.~\ref{fig:hqvs}(b). 
The two Majorana fermions in the two non-Abelian HQVs are in opposite spin sectors, and have different structures in their phase winding.

{\it Summary.}---
We have found two-fold non-Abelian anyons in a \tPt nematic superfluid, that is, non-Abelian HQVs 
characterized by a non-Abelian first homotopy group and Majorana fermions present inside their cores.
The HQVs are stabilized in the form of molecules through the interaction mediated by the uniaxial nematic component.
This is the first microscopic approach that describes the stability of HQVs, and we have clarified a new stabilization mechanism of HQVs due to the gap functions with a strong spin-orbit coupling.
Our finding will open a new era for 
non-Abelian anyons, possibly 
applicable to new directions in topological quantum computation
and neutron star physics.

\begin{acknowledgments}
{\it Acknowledgments.}---
Y.~M. thanks Robert Regan for useful comments. 
This work was supported by a Grant-in-Aid for Scientific Research on Innovative Areas ``Quantum Liquid Crystals (JP20H05163)'' from JSPS of Japan, and JSPS KAKENHI (Grant 
Nos.~JP18H01217, JP19K14662,
JP20K03860, JP20H01857, and JP21H01039). 
\end{acknowledgments}

\bibliographystyle{apsrev4-1}

\bibliography{reference}

\end{document}

\preprint{aps/}
\title{Supplemental Materials for ``Non-Abelian Half Quantum Vortices in $^{3}P_{2}$ Topological Superfluids''}

\author{Yusuke Masaki}
\email{MASAKI.Yusuke@nims.go.jp}
\affiliation{Research and Education Center for Natural Sciences, Keio University, Hiyoshi 4-1-1,
Yokohama, Kanagawa 223-8521, Japan}
\affiliation{Department of Physics, Tohoku University, Sendai, Miyagi 980-8578, Japan}
\affiliation{International center for Materials Nanoarchitectonics, National Institute for Materials Science, Tsukuba, Ibaraki 305-0044, Japan}
\author{Takeshi Mizushima}
\affiliation{Department of Materials Engineering Science, Osaka University, Toyonaka, Osaka 560-8531, Japan}
\author{Muneto Nitta}
\affiliation{Research and Education Center for Natural Sciences, Keio University, Hiyoshi 4-1-1,
Yokohama, Kanagawa 223-8521, Japan}
\affiliation{Department of Physics, Keio University, Hiyoshi 4-1-1, Japan}

\date{\today}

%%%%%%%%%%%%%%%%%%%%%%%%%%%%%%%%
\maketitle
%%%%%%%%%%%%%%%%%%%%%%%%%%%%%%%%
%%%%%%%%%%%%%%%%%%%%%%%%%%%%%%%%%%%%%%%%%%%%%%%%%%%%%%%%%%%%%%%%%%%%%%
This supplemental material consists of a summary of our microscopic approach, and boundary conditions in Secs.~S1 and S2, respectively. The order parameter profiles of two isolated half-quantum vortices (HQVs) are shown in Sec.~S3, and in Sec.~S4 mass current and spin current profiles are shown for a pair of such two HQVs. In Sec.~S5 we also discuss the symmetry and topology of zero-energy states bound to non-Abelian HQVs in the dihedral-four $D_{4}$ biaxial nematic (BN) phase of $^{3}P_{2}$ superfluids.

\section{S1. Microscopic theory}
In this section, we derive a quasiclassical framework and the Bogoliubov--de Gennes (BdG) equation from an interacting neutron system via a zero-range attractive ${}^{3}P_{2}$ force. The derivation can also be referred to our previous paper~\cite{Masaki:2019rsz}. The starting microscopic Hamiltonian $H$ is composed of the one-body terms $H_{1}$ and the interaction term $H_{2}$ given, respectively, in the following forms:
\begin{align}
H_{1} = \int \dd \bm{r} \sum_{\sigma,\sigma^{\prime}=\uparrow,\downarrow}
\psi_{\sigma}^{\dagger}(\bm{r})[h_{0}(-\ii \bm{\nabla})\delta_{\sigma,\sigma^{\prime}} + U_{\sigma\sigma^{\prime}}]\psi_{\sigma^{\prime}}(\bm{r}),~~~~~
H_{2} = - \int \dd \bm{r} \sum_{\alpha\beta= x,y,z}\dfrac{g}{2}T_{\alpha\beta}^{\dagger}(\bm{r})T_{\alpha\beta}(\bm{r}),
\end{align}
where $H_{1}$ consists of the kinetic energy $h_{0}(-\ii \bm{\nabla}) = (-\nabla^{2}/2m - \mu)$ measured from the chemical potential $\mu$, and the Zeeman energy $\hat{U}(\bm{r}) = -V_{\mathrm{Z}}\hat{\sigma}_{z}$ along the $z$ axis; $\hat{\cdot}$ denotes a 2 $\times$ 2 matrix in the spin space. The $^{3}P_{2}$ force with coupling strength $g > 0$ is represented by $H_{2}$, and the pair creation and annihilation operators $T^{\dagger}$ and $T$ are defined, respectively, by
\begin{align}
\hat{T}_{\alpha\beta}^{\dagger}(\bm{r}) = \sum_{\sigma,\sigma^{\prime}}
\psi_{\sigma}^{\dagger}(\bm{r})[t_{\alpha\beta,\sigma\sigma^{\prime}}^{*} (\ii \bar{\bm{\nabla}})\psi_{\sigma^{\prime}}^{\dagger}(\bm{r})],~~\text{and}~~~
T_{\alpha\beta}(\bm{r}) = \sum_{\sigma\sigma^{\prime}}[t_{\alpha\beta,\sigma\sigma^{\prime}}(-\ii \bar{\bm{\nabla}})\psi_{\sigma^{\prime}}(\bm{r})]\psi_{\sigma}(\bm{r}).
\end{align}
We have introduced a spin-momentum coupling in the pair force via the 2 $\times$ 2 matrix in spin space $\hat{t}_{\alpha\beta}$ defined by
\begin{align}
\hat{t}_{\alpha\beta}(-\ii \bar{\bm{\nabla}})
=\hat{t}_{\alpha\beta}(-\ii \bm{\nabla}/k_{\F})
=\ii \hat{\sigma}_{y}\left\{\dfrac{1}{2\sqrt{2}}\left[
\hat{\sigma}_{\alpha}(-\ii \bar{\nabla}_{\beta})
+ \hat{\sigma}_{\beta}(-\ii\bar{\nabla}_{\alpha})\right]
-\dfrac{1}{3\sqrt{2}}\delta_{\alpha\beta}\hat{\bm{\sigma}}\cdot(-\ii \bar{\bm{\nabla}})
\right\}
=[\hat{t}_{\alpha\beta}^{*}(\ii\bar{\bm{\nabla}})]^{*}.
\end{align}
Following Ref.~\cite{Masaki:2019rsz}, we obtain the quasiclassical transport equation known as the Eilenberger equation:
\begin{align}
0 = \ii \bm{v}_{\F} \cdot \bm{\nabla}\check{g} + [\ii \omega_{n}\check{\tau}_{z} + \check{u} + \check{\sigma}_{\Delta},\check{g}],
\end{align}
where $\check{\cdot}$ is a 4 $\times$ 4 matrix in the spin and Nambu space, $\check{\tau}_{z} = \mathrm{diag}(1,1, -1, -1)$ is the third component of the  Pauli matrices in the Nambu space, $\check{u} = V_{\mathrm{Z}}\mathrm{diag}(\hat{\sigma}_{z},\hat{\sigma}_{z}^{\mathrm{tr}})$ denotes the Zeeman field. The quasiclassical propagator $\check{g}$ is defined by
\begin{align}
\check{g}(\bm{k}_{\F},\bm{R};\ii\omega_{n}) 
&= \begin{bmatrix}\hat{g} & \hat{f} \\ -\hat{\bar{f}} & \hat{\bar{g}}\end{bmatrix}
= \oint_{C_{\mathrm{qc}}}\dfrac{\dd\xi}{ \ii \pi}\check{G}(\bm{k},\bm{R};\ii \omega_{n}),\\
\check{G}(\bm{k},\bm{R};\ii\omega_{n}) 
&= \int_{0}^{\beta}\dd \tau e^{\ii\omega_{n} \tau} 
\int \dd (\bm{r}_{1} - \bm{r}_{2})e^{-\ii \bm{k}\cdot(\bm{r}_{1} - \bm{r}_{2})}
\check{\tau}_{z}\braket{\mathcal{T}_{\tau}\vec{\Psi}(\bm{r}_{1},\tau)\vec{\Psi}^{\dagger}(\bm{r}_{2})}.
\end{align}
In the first line, the integral with respect to the energy $\xi = k^{2}/2m - \mu$ is performed along the contour $C_{\mathrm{qc}}$, which consists of the counterclockwise contour in the upper half $\xi$ plane and the clockwise contour in the lower half $\xi$ plane. Here, $\bm{R} = (\bm{r}_{1} + \bm{r}_{2})/2$ is a coordinate to describe the spatial inhomogeneity of the superfluid order parameter,     
$\mathcal{T}_{\tau}$ is an imaginary time ordered operator, and the Nambu spinor $\vec{\Psi}$ has been introduced as $\vec{\Psi}(\bm{r},\tau)
=(\psi_{\uparrow}(\bm{r},\tau),\psi_{\downarrow}(\bm{r},\tau),\psi_{\uparrow}^{\dagger}(\bm{r},\tau),\psi_{\downarrow}^{\dagger}(\bm{r},\tau))$ with $\psi_{\sigma}^{(\dagger)}(\bm{r},\tau) = e^{H\tau}\psi_{\sigma}^{(\dagger)}(\bm{r})e^{-H\tau}$. The order parameter matrix is defined by
\begin{align}
\check{\sigma}_{\Delta}(\bm{k}_{\F},\bm{R}) =  \begin{bmatrix}\hat{0} & \hat{\Delta} \\  -\hat{\Delta}^{\dagger} & \hat{0}\end{bmatrix},~~~
\hat{\Delta}(\bm{k}_{\F},\bm{R})
= \sum_{\alpha\beta}\mathcal{A}_{\alpha\beta}(\bm{R})\hat{\sigma}_{\alpha}\ii \hat{\sigma}_{y}\bar{k}_{\F,\beta},
\end{align}
where $\mathcal{A}_{\alpha\beta}$ is a 3 $\times$ 3 order parameter tensor of a spin-triplet $p$-wave superfluid, and $\bar{\bm{k}}_{\F} \equiv \bm{k}_{\F}/k_{\F}$ is a normalized Fermi momentum parametrized as $(\cos\alpha\sin\chi, \sin\alpha\sin\chi, \cos\chi)$. The order parameter tensor $\mathcal{A}$ is connected to the anomalous quasiclassical propagator $\hat{f}$ by the gap equation
\begin{align}
\mathcal{A}_{\alpha\beta} 
&= -\dfrac{\Delta_{\alpha\beta}(\bm{R}) + \Delta_{\beta\alpha}(\bm{R})}{2\sqrt{2}} + \dfrac{\sum_{\gamma}\Delta_{\gamma\gamma}(\bm{R})}{3\sqrt{2}}, \\
\Delta_{\alpha\beta}(\bm{R}) &= g\nu_{\mathrm{n}} \ii \pi T\sum_{n: |\omega_{n}| \le \omega_{\mathrm{c}}}\braket{\mathrm{Tr} \hat{t}_{\alpha\beta}(\bar{\bm{k}}_{\F})\hat{f}(\bm{k}_{\F},\bm{R};\ii\omega_{n})}_{\F},
\end{align}
where $g$ is the strength of the coupling constant and $\nu_{\mathrm{n}}$ is the density of state for the normal state at the Fermi energy, and $\braket{\cdots}_{\F}$ denotes the Fermi surface average defined by $\int \dd \alpha \int \dd \cos \chi/(4\pi) [\cdots]$. The cut-off energy of the Matsubara summation is denoted by $\omega_{\mathrm{c}}$ and related to $g\nu_{\mathrm{n}}$ as
\begin{align}
\dfrac{3}{g\nu_{\mathrm{n}}} = \log \dfrac{T}{T_{\mathrm{c}}} +  \sum_{n: |\omega_{n}| < \omega_{\mathrm{c}}}\dfrac{\pi}{\omega_{n}}.
\end{align}
An order parameter profile with a vortex is determined by 
self-consistently  solving the Eilenberger equation and the gap equations with the boundary condition explained below. In order to study the interaction energy of the HQVs, we utilize the Luttinger--Ward energy functional: 
\begin{align}
\mathcal{J}_{\mathrm{sn}}
= \dfrac{\nu_{\mathrm{n}}}{2}
\int_{0}^{1}\dd \lambda \mathrm{Sp}\left[\check{\sigma}_{\Delta}\left(\check{g}_{\lambda} - \dfrac{1}{2}\check{g}\right)\right],
\end{align}
where $\mathrm{Sp}[\cdots] = \ii \pi T \sum_{n} \int \dd \bm{R} \braket{\mathrm{Tr}[\cdots]}_{\F}$, and $\check{g}_{\lambda}$ is the solution to the equation $[\ii \omega_{n} \check{\tau}_{z} + \check{u} + \lambda\check{\sigma}_{\Delta},\check{g}_{\lambda}] + \ii \bm{v}_{\F}\cdot\bm{\nabla}\check{g}_{\lambda} = 0$. 

The Bogoliubov--de Gennes (BdG) equation also can be derived following Ref.~\cite{Masaki:2019rsz} as,
\begin{align}
&\check{\mathcal{H}}_{\BdG}(\bm{R})\vec{u}_{\nu}(\bm{R}) = \epsilon_{\nu} \vec{u}_{\nu}(\bm{R}),\\
&\check{\mathcal{H}}_{\BdG}(\bm{R}) =
\begin{pmatrix}
h_{0}(-\ii \bm{\nabla})\hat{1} + \hat{U} & \hat{\Delta}(\bm{R}) \\
-\hat{\Delta}^{*}(\bm{R}) & -h_{0}(\ii \bm{\nabla})\hat{1} - \hat{U}^{\mathrm{tr}}
\end{pmatrix},
\label{eq:bdg}
\end{align}
where $\hat{1}$ is the 2 $\times$ 2 identity matrix, and the gap matrix is given, using the anticommutator $\{a,b\}_{+} = ab + ba$, by
\begin{align}
\hat{\Delta}(\bm{R}) = \dfrac{\hat{\sigma}_{\alpha}\hat{\sigma}_{y}}{2k_{\F}}\left\{\mathcal{A}_{\alpha\beta}(\bm{R}),\nabla_{\beta}\right\}_{+}.
\end{align}

\section{S2. Boundary Condition and Basis Tensor of $\mathcal{A}$}
\label{sec:bc}

In this section, we explain in detail the boundary conditions for isolated HQVs and a pair of HQVs. 
For this purpose, it is convenient to expand the order parameter tensor $\mathcal{A}_{\alpha\beta}$ with respect to the $z$-component of the total angular momentum as $\mathcal{A}_{\alpha\beta}(\bm{R}) = \sum_{M = -2}^{2}\gamma_{M}(\bm{R})\Gamma_{M,\alpha\beta}$, where 
$\Gamma_{M}$ is the 3 $\times$ 3 basis tensor with the angular momentum $M$ such that $J_{z} \Gamma_{M} = M \Gamma_{M}$, and $\gamma_{M}(\bm{R})$ is a complex scalar function and denotes the projected order parameter onto the sector $M$. The operation $J_{z}$ on a tensor $\mathcal{A}$
is given by $[J_{z}\mathcal{A}]_{\alpha\beta} = \mathcal{A}_{\alpha\gamma}\ii\epsilon_{z \gamma \beta} - \ii \epsilon_{z\alpha\gamma} \mathcal{A}_{\gamma\beta}$ with antisymmetric tensor $\epsilon_{\alpha\beta\gamma}$. 
A representation of $\Gamma_{M}$ is given by
\begin{align}
\Gamma_{\pm2} = \dfrac{1}{2}
\begin{pmatrix}
1 & \pm \ii & 0 \\
\pm \ii & - 1 & 0\\
0 & 0 & 0
\end{pmatrix},~~
\Gamma_{\pm1} = \dfrac{1}{2}
\begin{pmatrix}
0 & 0 & \mp 1\\
0 & 0 & -\ii \\
\mp 1 & -\ii & 0
\end{pmatrix},~~
\Gamma_{0} = \dfrac{1}{\sqrt{6}}
\begin{pmatrix}
-1 & 0 & 0\\
0 & -1 & 0 \\
0 & 0 & 2
\end{pmatrix}.\end{align}
The $D_{4}$-BN state with point nodes along the $z$ axis is represented for this basis set as $|\gamma_{2}| = |\gamma_{-2}|$ and $\gamma_{0,\pm1} = 0$. In the following, we consider asymptotic forms of vortices, which give boundary conditions for the above equation set.
An integer vortex for such a $D_{4}$-BN state has an asymptotic form given by
\begin{align}
\mathcal{A}_{\alpha\beta}(\theta) = \sum_{M=\pm2}e^{ \ii \kappa \theta}\gamma_{M}\Gamma_{M,\alpha\beta}~~(\kappa = \pm1).
\end{align}
This boundary condition breaks axial symmetry 
which is described by the form $\mathcal{A} = \sum_{M}e^{\ii (\kappa - M)\theta}\gamma_{M}\Gamma_{M,\alpha\beta}$. 
In addition to such a non-axisymmetric integer vortex, an HQV is topologically allowed in the $D_{4}$-BN state. Its asymptotic form is given by
\begin{align}
\mathcal{A}_{\alpha\beta}(\theta) = \sum_{M=\pm2}e^{ \ii (\kappa - n M) \theta}\gamma_{M}\Gamma_{M,\alpha\beta}~~(\kappa = \pm 1/2, n = \pm1/4),  \label{eq:bc-hqv}
\end{align}
which indicates that there are two kinds of HQVs in terms of the relative sign between $\kappa$ and $n$. Without loss of generality, we set $\kappa = 1/2$ in the main text and in the following. It is proposed that an integer vortex with $\kappa = 1$ can split into two HQVs with $(\kappa,n) = (1/2, 1/4)$ and $(\kappa,n) = (1/2, -1/4)$~\cite{Masuda:2016vak}. Such a configuration is schematically given as follows: Let a vortex center with $n=-1/4$ ($n=1/4$) be $\bm{R}_{1}$ $(\bm{R}_{2})$, and define the polar angle measured from its center as $\theta_{1}$ ($\theta_{2}$). An asymptotic form of two HQVs which is equivalent to that of a single integer vortex is given by
\begin{align}
\mathcal{A}(\theta_{1},\theta_{2}) = \sum_{M=\pm2} e^{\ii (\kappa + nM)\theta_{1} + \ii (\kappa - nM)\theta_{2}}\gamma_{M}\Gamma_{M,\alpha\beta}~~(\kappa = 1/2, n = 1/4), \label{eq:bc-twohqvs}
\end{align}
 where the phase behaves as $2 \kappa \theta$ for $|\bm{R}| \to \infty (\theta_{1,2} \to \theta)$. 

\begin{figure}
\includegraphics[width=\hsize]{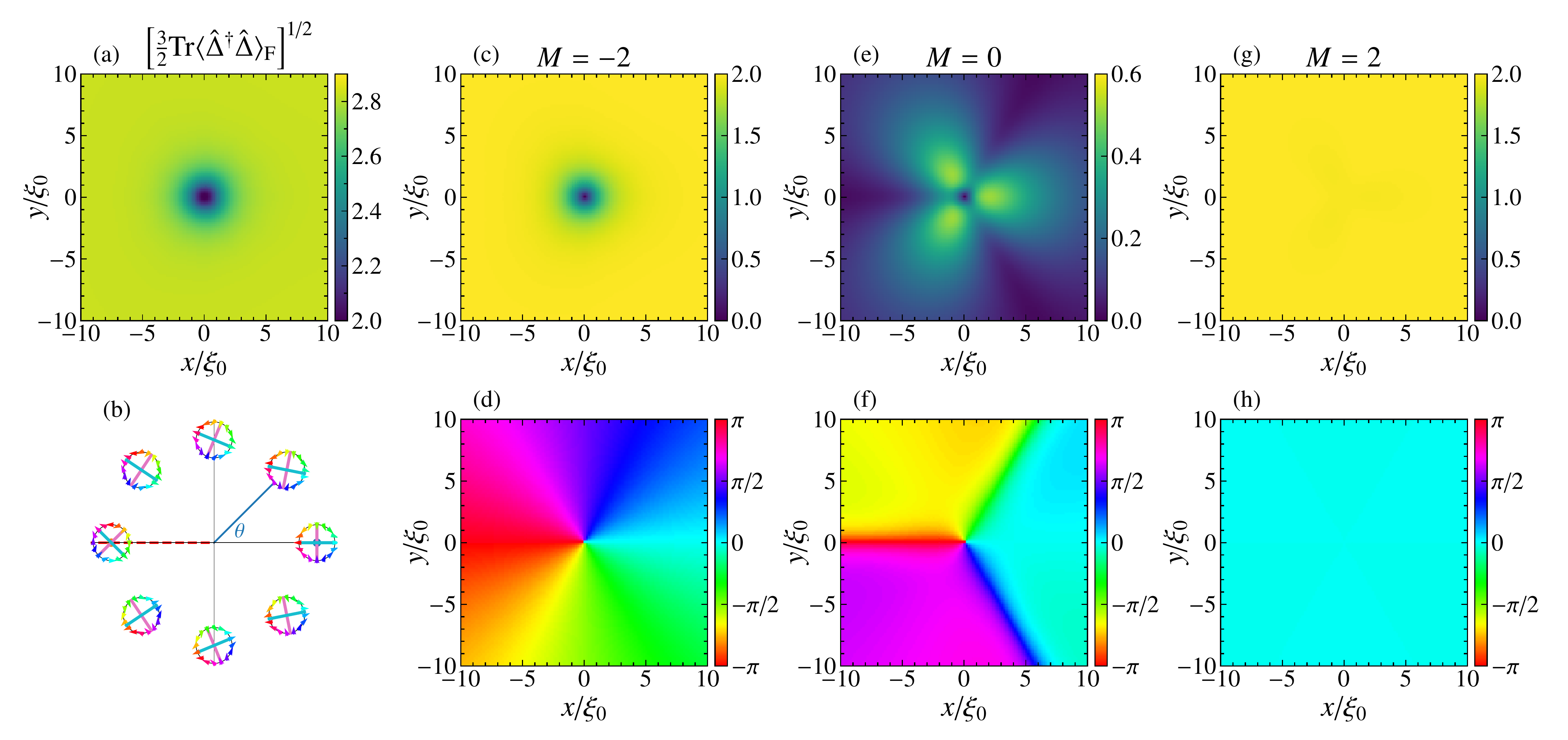}
\caption{HQV with $(\kappa,n) =(1/2, -1/4)$. 
	     (a) Total intensities $[\sum_{M}|\gamma_{M}|^{2}]^{1/2}$. 
	     (b) Schematic image of the boundary condition. 
	     The amplitudes $|\gamma_{M}|$ for $M=-2$, $0$, and $2$ are shown in (c), (e), and (g), respectively.
	     The phases $\tan^{-1}(\gamma_{M})$ for $M=-2$, $0$, and $2$ are shown in (d), (f), and (h), respectively.
	 	}
\label{fig:-1/4}
\end{figure}
\begin{figure}
\includegraphics[width=\hsize]{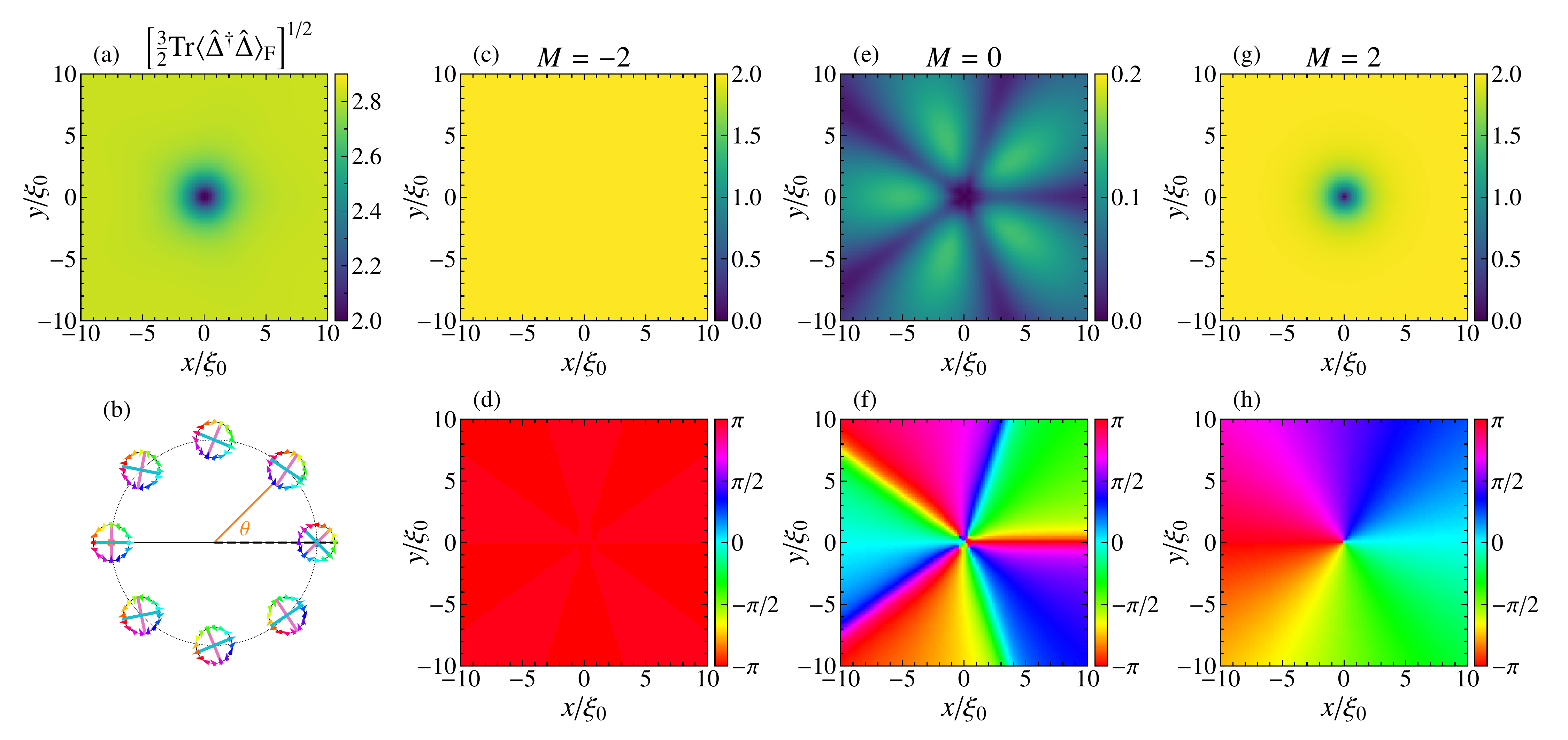}
\caption{HQV with $(\kappa,n) =(1/2, 1/4)$. 
	     (a) Total intensities $[\sum_{M}|\gamma_{M}|^{2}]^{1/2}$. 
	     (b) Schematic image of the boundary condition. 
	     (c), (e), (g) Amplitudes $|\gamma_{M}|$ for $M = -2$, $0$ ,and $2$.
	     (d), (f), (h) Phases $\tan^{-1}(\gamma_{M})$ for $M = -2$, $0$ ,and $2$.}
\label{fig:1/4}
\end{figure}

\section{S3. Isolated Half Quantum Vortices}
\label{sec:hqv}

In this section, we present self-consistent solutions of isolated HQVs. As mentioned in the previous section and in the main text, 
there are two kinds of HQVs characterized by $n = \pm 1/4$ implying two possible rotational directions of the triad relative to the vorticity. 
In Fig.~\ref{fig:-1/4}, a HQV solution with $(n,\kappa) = (1/2, -1/4)$ is shown. The relative phase of $\gamma_{2}$ and $\gamma_{-2}$ is set to zero in Eq.~\eqref{eq:bc-hqv}, and the schematic image of the boundary condition is shown in panel (b). In panel (b), $\theta$ denotes the azimuthal angle in  real space and the red dashed line implies the branch cut. The colored arrows in each circular object represent the direction of the $d$ vectors, whose color bar is identical to those of panels (b), (f), and (h). The cyan and the magenta lines in each object stand for the directions of the eigenvalues $1$ and $-1$ of $\mathcal{A}_{\alpha\beta}(\bm{R})$, respectively. Panel (a) shows the total amplitude defined by $\{\frac{3}{2}\mathrm{Tr}\braket{[\hat{\Delta}(\bm{k}_{\F},\bm{R})]^{\dagger}\hat{\Delta}(\bm{k}_{\F},\bm{R})}_{\F}\}^{1/2} = [\sum_{M}|\gamma_{M}(\bm{R})|^{2}]^{1/2}$. Panels (c), (e), and (g) show $|\gamma_{-2}|$, $|\gamma_{0}|$, and $|\gamma_{2}|$, respectively, while panels (d), (f), and (h) show the corresponding phase $\arg (\gamma_{M})$. As expected from the boundary condition, the sector $M = -2$ has single winding, while the sector $M = 2$ has no winding, which results in an almost uniform amplitude profile. It is remarkable that the amplitude of the induced component $M=0$ seems three-fold. It has single phase winding around the core, but opposite to the winding in $M=-2$. Although the phase change along the loop around the core is non linear, the opposite single winding can be understood as an analogy to a vortex in a spinless chiral $p$-wave superconductor. The gap matrices contributed from the sectors $M = - 2$ and $M=0$ are given, respectively, by
\begin{align}
\hat{\Delta}_{-2} = \gamma_{-2}(\bm{R})
\begin{pmatrix}
0 & 0 \\
0 & \bar{k}_{x} - \ii \bar{k}_{y}
\end{pmatrix},~~
\hat{\Delta}_{0} = \dfrac{\gamma_{0}(\bm{R})}{\sqrt{6}}
\begin{pmatrix}
\bar{k}_{x} - \ii \bar{k}_{y} & 0 \\
0 & - (\bar{k}_{x} + \ii \bar{k}_{y})
\end{pmatrix}, \label{eq:gap-matrix:-1/4}
\end{align}
where we set $\bar{k}_{z} = 0$ for simplicity. In the case of an axisymmetric vortex in a spinless chiral $p$-wave superconductor, 
the induced component has the opposite chirality relative to the component which has a non-zero order parameter far from the vortex core~\cite{Heeb1999,MatsumotoHeeb2001}. Here we call the latter component the {\it bulk component}. In this case, the vorticity of the induced component is determined such that a simultaneous rotation in the real space and the momentum space gives the same phase change between the bulk component and the induced component. In more details, let $\kappa_{\mathrm{b}}$ ($\kappa_{\mathrm{i}}$) and $\chi_{\mathrm{b}}$ ($\chi_{\mathrm{i}}$) be the vorticity and  chirality of the bulk component (the induced component), respectively. The chirality is defined so that $\bar{k}_{x} + \ii \chi \bar{k}_{y} = e^{\ii \chi \alpha}$, and it takes either $1$ or $-1$, and $\chi_{\mathrm{i}} = -\chi_{\mathrm{b}}$. The phase factor is given by $\kappa_{\mathrm{b}} \theta + \chi_{\mathrm{b}}\alpha$ and $\kappa_{\mathrm{i}} \theta + \chi_{\mathrm{i}}\alpha$, and the simultaneous rotation by angle $\Delta \theta$ gives the phase change $(\kappa_{\mathrm{b}} + \chi_{\mathrm{b}})\Delta \theta$ and $(\kappa_{\mathrm{i}} + \chi_{\mathrm{i}})\Delta \theta$, respectively. Therefore, the vorticity for the induced component is determined as 
$\kappa_{\mathrm{i}} = \kappa_{\mathrm{b}} + \chi_{\mathrm{b}} - \chi_{\mathrm{i}} = \kappa_{\mathrm{b}} + 2 \chi_{\mathrm{b}} \equiv (\kappa_{\mathrm{b}} + \chi_{\mathrm{b}}) + \chi_{\mathrm{b}}$, where $\kappa_{\mathrm{b}} + \chi_{\mathrm{b}}$ is the sum of the vorticity and  chirality, characterizing two inequivalent vortices in chiral $p$-wave superconductors.
By applying this argument to the present HQV case, the winding of the induced component is obtained as follows: From Eqs.~\eqref{eq:gap-matrix:-1/4} and \eqref{eq:bc-hqv}, $\kappa_{\mathrm{b}} = 1$ and $\chi_{\mathrm{b}} = -1$ for $M = -2$, and thus we obtain $\kappa_{\mathrm{i}} = - 1$. This is consistent with the opposite single winding indicated by panel (f). Again we remark that the phase change is non linear and the amplitude does not retain the axial symmetry in the present case. In terms of the spin sectors, the winding in the bulk component appears only in the spin down sector as seen from Eq.~\eqref{eq:gap-matrix:-1/4}.
 
Next, we present a self-consistent solution of the other isolated HQV. In order to describe the HQV in the left hand side of a pair of the HQVs, we shift the phase of $\gamma_{-2}$ by $\pi$ in Eq.~\eqref{eq:bc-hqv}, as seen from Eq.~\eqref{eq:bc-twohqvs} by setting $\theta_{1} = \pi$ for $\bm{R}_{2} = (0,0)$ and $\bm{R}_{1}=(\infty, 0)$. The numerical solution of this HQV is shown in Fig.~\ref{fig:1/4}. The meaning of each panel is the same as in Fig.~\ref{fig:-1/4}. The boundary condition schematically shown in panel (b) stands for the opposite rotation compared with Fig.~\ref{fig:-1/4}(b).
Correspondingly, the sector with $M = -2$ has no winding, while that with $M=2$ has a single winding. The gap matrix contributed from the sector $M=2$ is written down as 
\begin{align}
\hat{\Delta}_{2} = -\gamma_{2}(\bm{R})
\begin{pmatrix}
\bar{k}_{x} + \ii \bar{k}_{y} & 0 \\  0 & 0
\end{pmatrix}.\label{eq:gap-matrix:1/4}
\end{align}
It can be seen that only the spin-up sector in the bulk component has the winding in this case,
in contrast to Eq.~(\ref{eq:gap-matrix:-1/4}) for the spin-down sector in the previous case.
 The winding of the induced component can be understood by applying the above argument. In the present case, it follows from Eqs.~\eqref{eq:bc-hqv} and \eqref{eq:gap-matrix:1/4} that $\kappa_{\mathrm{b}} = 1$ and $\chi_{\mathrm{b}}= 1$ for $M=2$, while $\chi_{\mathrm{i}} = -1$ for the spin-up sector of $M=0$ as seen from Eq.~\eqref{eq:gap-matrix:-1/4}.
Therefore, we obtain $\kappa_{\mathrm{i}} = 3$, which is consistent with the winding indicated by Fig.~\ref{fig:1/4}(f). We note that its change is nonlinear as well as the previous HQV. Correspondingly, the amplitude $|\gamma_{0}|$ is five fold, 
in contrast to a three-fold symmetry in the previous case.

\section{S4. Mass Current and Spin Current for a Pair of HQVs}
\begin{figure}
\includegraphics[width=\hsize]{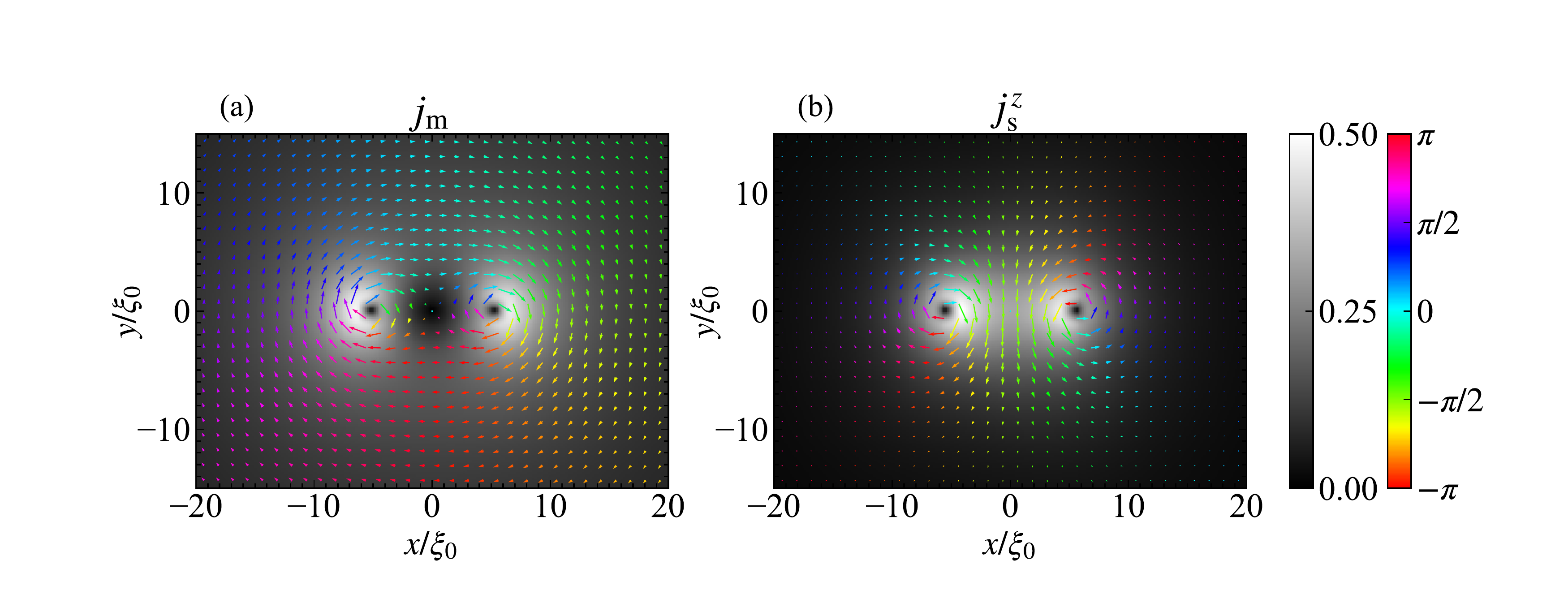}
\caption{(a) Mass current $j_{\mathrm{m}}$ and (b) spin current $j_{\mathrm{s}}^{z}$ for the $z$ component for a pair of HQVs. 
            In each panel, the direction of the in-plane current is indicated by the colors and the directions of the arrows, 
            while its intensity is indicated by the lengths of the arrows and the colormap behind the arrows.
            }
\label{fig:current}
\end{figure}
In this section, we show the spatial profiles of the mass current density $j_{\mathrm{m}}$ and the spin current density with spin along the $z$ axis $j_{\mathrm{s}}^{z}$ for a pair of HQVs.
Within the quasiclassical theory, these currents are calculated by
\begin{align}
\bm{j}_{\mathrm{m}}(\bm{R}) 
&= \dfrac{\ii \pi T \nu_{\mathrm{n}}}{2m}\sum_{n:|\omega_{n}|\le \omega_{\mathrm{c}}}\mathrm{Tr}\left[\braket{{\bm{k}}_{\F}\check{\tau}_{z}\check{g}(\bm{k}_{\F},\bm{R};\ii\omega_{n})}_{\F}\right],
\\
j_{\mathrm{s}}^{z}(\bm{R})
&= \dfrac{\ii \pi T \nu_{\mathrm{n}}}{2m}\sum_{n:|\omega_{n}|\le \omega_{\mathrm{c}}}\mathrm{Tr}\left[\braket{{\bm{k}}_{\F}\check{\tau}_{z}\check{g}(\bm{k}_{\F},\bm{R};\ii\omega_{n})\check{S}_{z}}_{\F}\right],
\end{align}
where $\check{S}_{\mu = x,y,z} = \mathrm{diag}(\hat{\sigma}_{\mu},\hat{\sigma}_{\mu}^{\mathrm{tr}})/2$. The summations over the Matsubara frequency are cut-off by $\omega_{\mathrm{c}}$ which is used for the above gap equation, and it is (at least) qualitatively correct. In Fig.~\ref{fig:current}, we show the in-plane components of $j_{\mathrm{m}}$ in panel (a) and $j_{\mathrm{s}}^{z}$ in panel (b). The currents along the $z$ axis are zero, and the spin currents for $\mu = x$ and $y$ are zero as well. In each panel, the colored arrows show the direction of the current and its intensity, and the colormap behind the arrows also indicates the intensity of the current. The mass current reflects the vorticity of the two HQVs, and it rotates in the same sense around the two cores. In the region far from the core, it gradually decays in a way similar to that of a singly quantized vortex. Between the two cores, the mass currents contributed from the two cores cancel out. The spin current has an opposite behavior relative to the mass current: it rotates in the opposite sense between the cores because the localized states contributing to the current have opposite spins. As a result, the contributions far from the cores vanishes, while the region between the cores have rather strong flows. 

\section{S5. Symmetry and topology of vortex-bound states in the $D_{4}$-BN phase}

Here we discuss the symmetry and topology of zero-energy states bound at a molecule of non-Abelian HQVs in the $D_{4}$-BN state. For comparison, we also show the axisymmetric $o$-vortex in the $D_{4}$-BN state, which is the most symmetric vortex with normal core, hosting multiple Majorana zero modes. In Ref.~\cite{Masaki:2019rsz}, we demonstrated that under constraints of the axial symmetry of vortices, the $\varv$ vortex with the uniaxial-nematic superfluid core is thermodynamically stable in low magnetic fields, while the $o$ vortex becomes stable as the magnetic field increases. We also found that the $o$ vortex has topologically-protected zero-energy modes, while the axisymmetric $\varv$ vortex has a finite excitation gap in the low-lying quasiparticle spectrum. In this work, we investigate the stability and microscopic structure of non-Abelian HQVs without assuming axial symmetry. As mentioned in the main text, the low-lying quasiparticle structure in the HQV is distinct from those of the $o$ vortex and $\varv$ vortex, since a single zero energy state appears at each core of HQVs. Below, we uncover the robustness and non-Abelian anyonic nature of the zero-energy states bound at a molecule of non-Abelian HQVs on the basis of symmetry and topology.

\subsection{S5.1. Symmetry classification}

To unveil the symmetry and topology of vortices, following Refs.~\cite{tsutsumiPRB15,mizushimaJPSJ16,shiozakiPRB14}, we start with the semiclassical approximation where the Hamiltonian varies slowly in the real-space coordinate system. The Bogoliubov quasiparticle structure is governed by the BdG Hamiltonian in Eq.~\eqref{eq:bdg}. The $^{3}P_{2}$ order parameter in the BdG Hamiltonian has the spatial modulation due to a vortex line. In the semiclassical approximation, the spatial modulation is considered as adiabatic changes as a function of the real-space coordinate surrounding the defect with an angle $\theta$, where we introduce the cylindrical coordinate ${\bm R}=(\rho,\theta,z)$. Then, the BdG Hamiltonian in the base space, $(\bm{k},\theta)$ is obtained from Eq.~\eqref{eq:bdg}, as 
\begin{align}
\check{\mathcal{H}}({\bm k},\theta) 
= \begin{pmatrix}
{h}_0({\bm k})\hat{1}+\hat{U} & \hat{\Delta}({\bm k},\theta) \\
\hat{\Delta}^{\dag}({\bm k},\theta) & -h_{0}({\bm k})\hat{1} - \hat{U}^{\mathrm{tr}}
\end{pmatrix},
\label{eq:bdg2}
\end{align}
where $\hat{U}$ is the Zeeman energy due to a magnetic field ${\bm H}$, and the $2\times 2$ spin matrix of the $^{3}P_{2}$ pair potential is given by 
\begin{align}
\hat{\Delta}({\bm k},\theta) = i \hat{\sigma}_{\mu}\mathcal{A}_{\mu i}(\theta) k_i/k_{\F} \hat{\sigma}_{y}.
\end{align}
Here we consider the $D_{4}$-BN phase of  $^{3}P_{2}$ superfluids. The asymptotic form of the order parameter with an isolated vortex satisfies the boundary condition
\begin{align}
\mathcal{A}_{\mu i}(\theta) = \Delta_0 e^{i\kappa\theta}R_n(\theta)
\begin{pmatrix}
1 & 0 & 0 \\ 0 & -1 & 0 \\ 0 & 0 & 0 
\end{pmatrix}R^{\mathrm{tr}}_{n}(\theta),
\label{eq:bc}
\end{align}
where $R^{(z)}_n(\theta)\in \mathrm{SO}(2)$ is the rotation matrix about $z$ axis by angle $n\theta$
\begin{align}
R^{(z)}_n(\theta) = \begin{pmatrix}
\cos(n\theta)  & -\sin(n\theta) & 0 \\ \sin(n\theta) & \cos(n\theta) & 0 \\ 0 & 0 & 1
\end{pmatrix}.
\end{align}
We assume the spatial uniformity of the order parameter along the vortex line, \ie, the $z$ axis. The parameters $\kappa$ and $n$ in Eq.~\eqref{eq:bc} represents the $\mathrm{U}(1)$ phase winding and the $n$-fold texture of the $D_{4}$-BN order along the azimuthal angle $\theta$, respectively (see Fig.~1 in the main text).

%-------------------------------------------------------
\begin{table*}[t!]
\caption{\label{table1}Classification of vortices for the $D_{4}$-BN phase in terms of the vorticity $\kappa$, the $n$-fold rotation of the $D_{4}$-BN order, the mirror ($M_{xy}$) and $P_{3}$ (magnetic $\pi$ rotation) symmetries, and topological invariants $\mathrm{Ch}_{2}\in 2\mathbb{Z}$, $\nu_{\lambda}\in\mathbb{Z}_{2}$, and $w_{\mathrm{1d}}\in\mathbb{Z}$, where $\lambda=\pm i$ is the eigenvalue of $M_{xy}$. The column ``Core states'' corresponds to the superfluid components which may be induced around the vortex core (also see the discussion below Eq.~\eqref{eq:p3tr}), and ``non-Abelian'' is the existence of the zero-energy state (ZES) which has the nature of a non-Abelian anyon. We numerically confirmed the self-consistent solutions of axisymmetric $o$- and $\varv$-vortices in addition to HQVs, while no stable solutions of $u$, $w$, and $u \varv w$ vortices have been found with assuming  the axial symmetry.} 
\begin{ruledtabular}
%\centering
\begin{tabular}{cccccccccccccc}
& & Symmetry & $M_{xy}$ & $\kappa$ & $n$ & Core states & $\mathrm{Ch}_{2}$ & $(\nu_{\lambda=+i},\nu_{\lambda=-i})$ & $|w_{\mathrm{1d}}|$ & ZES & Non-Abelian anyon \\
\hline
&$o$ vortex & $P_{1},P_{2},P_{3}$  & $\surd$   & $\mathbb{Z}$ & $\mathbb{Z}$ & Re$\gamma_{0}$ & 0 & $(1,-1)$ & 2 & Yes & Yes  & \\
&$u$ vortex & $P_{1}$ & --- & $\mathbb{Z}$ & $\mathbb{Z}$ & $\gamma_0$  &   &  &  &  & & \\
&$\varv$ vortex & $P_{2}$ & --- & $\mathbb{Z}$ & $\mathbb{Z}$ & $\mathrm{Re}\gamma_{\pm 1}$, $\mathrm{Re}\gamma_0$  & 0  & --- & --- & No & & \\
&$w$ vortex & $P_{3}$ & --- & $\mathbb{Z}$ & $\mathbb{Z}$ & $\mathrm{Im}\gamma_{\pm 1}$, $\mathrm{Re}\gamma_0$  &   &  &  & & & \\
&$u \varv w$ vortex & --- & --- & $\mathbb{Z}$ & $\mathbb{Z}$ & $\gamma_{0,\pm 1}$  &   &  &  &  &  & \\
& HQV       & $P_{3}$ & $\surd$ & $\mathbb{Z}+ 1/2$ & $\mathbb{Z}\pm 1/4$ & $\mathrm{Re}\gamma_0$ &  0 &  $(1,0)$ or $(0,-1)$ & 1 & Yes & Yes & \\
\end{tabular}
\end{ruledtabular}
\label{table:critical_exponents}
\end{table*}
%-------------------------------------------------------/

Let us consider systems invariant under a $\mathrm{U}(1)$ gauge transformation and an $\mathrm{SO}(3)$ spin-momentum rotation. The continuous symmetry group, $G=\mathrm{U}(1)\times \mathrm{SO}(3)$, acts on the the $3\times 3$ traceless symmetric tensor, $\mathcal{A}_{\mu i}$, as 
$\mathcal{A} \rightarrow e^{i\varphi}g\mathcal{A}g^{\mathrm{tr}}$, where $e^{i\varphi}\in \mathrm{U}(1)$ and $g\in\mathrm{SO}(3)$ are elements associated with a gauge transformation and simultaneous spin-orbit rotation, respectively. The $D_{4}$-BN order, $\mathcal{A}_{\mu i}^{\mathrm{BN}}$, is invariant under the dihedral-four ($D_{4}$) group, whose elements $(e^{i\alpha},g)$ are 
\begin{align}
D_{4} = \{ (1,{\bm 1}), (-1,R^{z}_{4}),  (1, I^{z}), (-1,I^{z}R_{4}^{z}), (1, I^{x}), (1, I^{y})\},
\label{eq:D4}
\end{align}
where $I^{j}$ and $R^z_4$ represent $\pi$ rotations around the $j$ axis ($j=x,y,z$) and $\pi/2$ ($n=4$) rotations around the $z$ axis, respectively, and ${\bm 1}$ is the $3\times 3$ unit matrix. The OP manifold in the $D_{4}$-BN state is then characterized by the broken symmetry, $R=[\mathrm{U}(1)\times \mathrm{SO}(3)]/D_{4}$. The topological charges of line defects are characterized by the first homotopy group 
\begin{align}
\pi_{1}(R) = \mathbb{Z}\times_{h} D_{4}^{\ast}
\end{align}
where $\times_{h}$ denotes a product 
defined in Ref.~\cite{Kobayashi:2011xb}.
This ensures two different classes of topological line defects: Vortices with commutative topological charges and vortices with non-commutative topological charges. The former includes the $o$ and $\varv$ vortices, while an example of the latter is the non-Abelian HQV. Integer vortices are characterized by $\kappa \in \mathbb{Z}$ and $n\in\mathbb{Z}$ in Eq.~\eqref{eq:bc}. The HQV has a half integer of vorticity, $\kappa \in \mathbb{Z}+1/2$, which is accompanied by the $\pi$ jump in the $\mathrm{U}(1)$ phase, $e^{\pm i\theta/2}$. In the $D_{4}$-BN state, however, as the order parameter tensor has the $D_{4}$ symmetry as in Eq.~\eqref{eq:D4}, 
$R^z_4 \mathcal{A}^{\mathrm{BN}} R^{z\mathrm{tr}}_4 = - \mathcal{A}^{\mathrm{BN}} $, 
the discontinuity is compensated by the 4-fold rotation with $n \in \mathbb{Z} \pm 1/4$. For $\kappa>0$, the choice of $n=+1/4$ ($-1/4$) corresponds to the clockwise (counterclockwise) texture of the $D_{4}$-BN gap structure along the azimuthal direction.

It is convenient to expand $\mathcal{A}_{\mu\nu}$ in terms of eigenstates of the total angular momentum $J_{z}$, $\Gamma_{M}$, as in Sec.~S2. In this basis, the $D_{4}$-BN order parameter is composed of the equal contributions of the $M=+2$ and $M=-2$ components, where the former (latter) corresponds to the Cooper pairing with angular momentum $L_{z}=+1$ ($-1$) and spin angular momentum $S_{z}=+1$ ($-1$). 
On the basis of $\Gamma_{M}$, the vortex state can be expressed as 
\begin{align}
\mathcal{A}_{\mu\nu}(\rho,\theta) = \sum_{M=-2}^{2} \gamma_{M}(\rho,\theta) \Gamma_{M,\mu\nu}.
\label{eq:ansatz}
\end{align}
Particularly, the asymptotic form of an isolated HQV at $\rho \rightarrow \infty$ is obtained from Eq.~\eqref{eq:bc} as
\begin{align}
\mathcal{A}_{\mu\nu}(\theta) = \sum_{M=-2,2} \gamma_{M} e^{i \kappa_{M} \theta} \Gamma_{M,\mu\nu}.
\end{align}
The winding number of Cooper pairs with $J_{z}=M$, $\kappa_{M}$, must be an integer, $\kappa_{M}\in\mathbb{Z}$, and is represented by $\kappa_{M} = \kappa \theta - \varphi$. An isolated HQV is characterized by the vorticity $\kappa = 1/2$ and the rotation angle of the triad $\varphi = n\theta = + \theta/4$ ($\varphi = -\theta/4$). 
The combinations of $(\kappa,n)$ in the other vortices are summarized in Table~\ref{table1}. On the basis of the eigenstate $\Gamma_{M}$, as mentioned above, the asymptotic form of the HQV is recast into the superposition of the $ L_{z}=S_{z}=+1$ Copper pairs with vorticity $\kappa _{M=+2} =0$ ($1$) and the $L_{z}=S_{z}=-1$ Copper pairs with vorticity $\kappa _{M=-2} =1$ ($0$), i.e., $(k_{x}+ik_{y})\ket{\uparrow\uparrow}+e^{i\theta}(k_{x}-ik_{y})\ket{\downarrow\downarrow}$ for $(\kappa,n)=(1/2,1/4)$. As mentioned in Sec.~S3 and the main text, however, the superfluid component $\gamma_{M=0}$ appears around the vortex core within the length scale of $\xi$.

Let us now summarize the symmetries of the BdG Hamiltonian in Eq.~\eqref{eq:bdg}. The BdG Hamiltonian with line defects in three-dimensional system belongs to the class D in the ten-fold way of the Altland--Zirnbauer symmetry classes~\cite{Schnyder:2008tya}. The Hamiltonian breaks time-reversal symmetry but holds  particle-hole symmetry
\begin{align}
\check{\mathcal{C}} \check{\mathcal{H}}({\bm k},\theta) \check{\mathcal{C}}^{-1} = -\check{\mathcal{H}}(-{\bm k},\theta),
\label{eq:phs}
\end{align}
where $\check{\mathcal{C}} = \check{\tau}_{x} K$ ($K$ is the complex conjugation operator). 

In addition to the particle-hole symmetry, three different types of discrete symmetries, $\{P_{1},P_{2},P_{3}\}$, are relevant to vortices. These symmetry operators are composed of discrete elements of the symmetry group $G$, such as the time-reversal operator $\hat{\mathcal{T}}=-i\hat{\sigma}_{y}K$, the spatial inversion ${P}$, the discrete phase rotation $U_{\pi}=e^{i(\kappa+1)\pi}$, and the joint $\pi$-rotation about an axis perpendicular to the vortex line (e.g., the $x$ axis), $\hat{C}_{2,x}=e^{-i\hat{J}_{x}\pi}$. The elements of such symmetry are given by 
\begin{align}
\hat{P}_{1}=PU_{\pi}\hat{1}, ~ \hat{P}_{3} = \hat{\mathcal{T}}\hat{C}_{2,x},
\end{align}
and $\hat{P}_{2}=\hat{P}_{1}\hat{P}_{3}$. For a vortex preserving these symmetries, the BdG Hamiltonian in Eq.~\eqref{eq:bdg2} obeys
\begin{align}
\check{\mathcal{P}}_{1} \check{\mathcal{H}}(k_{x},k_{y},k_{z},\theta) \check{\mathcal{P}}_{1}^{-1} &= \check{\mathcal{H}}(-k_{x},-k_{y},-k_{z},\theta+\pi), \\
\check{\mathcal{P}}_{3} \check{\mathcal{H}}(k_{x},k_{y},k_{z},\theta) \check{\mathcal{P}}_{3}^{-1} &= \check{\mathcal{H}}(-k_{x},k_{y},k_{z},-\theta)
\end{align}
where $\check{\mathcal{P}}_{3} \equiv \mathrm{diag}(\hat{P}_{1},\hat{P}_{1}^{\ast})$ and $\check{\mathcal{P}}_{3} \equiv \mathrm{diag}(\hat{P}_{3},\hat{P}_{3}^{\ast})=\mathrm{diag}(i \hat{\sigma}_{z}, -i \hat{\sigma}_{z})K$ 
 are the $P_{1,2,3}$ operators extended to the Nambu space. The $o$ vortex is the most symmetric vortex preserving all discrete symmetries, $\{P_{1},P_{2},P_{3}\}$. 
The $P_{3}$ operator composed of the $\pi$-rotation about the $x$ axis and $\mathcal{T}$ flips the momentum and spin as ${\bm k}\rightarrow (-k_{x},k_{y},k_{z})$ and ${\bm \sigma}\rightarrow (\sigma_{x},-\sigma_{y}-\sigma_{z})$ and changes the azimuthal angle $\theta\rightarrow -\theta$. The $^{3}P_{2}$ order parameter is transformed as
\begin{align}
P_{3}:~\mathcal{A}(\theta) \rightarrow \begin{pmatrix}
 \mathcal{A}^{\ast}_{xx}(-\theta) & -\mathcal{A}^{\ast}_{xy}(-\theta) & -\mathcal{A}^{\ast}_{xz}(-\theta) \\
-\mathcal{A}^{\ast}_{xy}(-\theta) &  \mathcal{A}^{\ast}_{yy}(-\theta) &  \mathcal{A}^{\ast}_{yz}(-\theta) \\
-\mathcal{A}^{\ast}_{xz}(-\theta) &  \mathcal{A}^{\ast}_{yz}(-\theta) &  \mathcal{A}^{\ast}_{zz}(-\theta) 
\end{pmatrix}, \label{eq:p3tr}
\end{align}
corresponding to $\gamma_{M}(\rho,\theta) \rightarrow (-1)^{M}\gamma^{\ast}_{M}(\rho,-\theta)$. In the $D_{4}$-BN state, an axisymmetric vortex represented by Eq.~\eqref{eq:ansatz} with $\gamma_{M}(\rho,\theta)=\gamma_{M}e^{i(\kappa - M )\theta}$ maintains the $P_{3}$ symmetry when $\mathrm{Re}\gamma_{\pm 1}(\rho)=0$. 
The $P_{1}$ symmetry requires the order parameters to satisfy the relations, $\gamma_{M}(\rho, \theta) = -\gamma_{M}(\rho,\theta+\pi)$. For axisymmetric vortices, this implies that nonvanishing $\gamma_{\pm 1}$ or an even $\kappa_{M}$ breaks the $P_{1}$ symmetry.
In the same manner, the $P_{2}$ symmetry leads to the relations $\gamma_{M}(\rho,\theta) \rightarrow -(-1)^{M}\gamma^{\ast}_{M}(\rho,-\theta+\pi)$. For axisymmetric vortices with an odd $\kappa$, the $P_{2} $ symmetry is preserved if $\mathrm{Im}\gamma_{M}(\rho)=0$.
The vortex state preserving only the $P_{2}$ ($P_{3}$) symmetry is referred to as the $\varv$-vortex ($w$-vortex) state. As discussed in Sec.~S3 and Figs.~S2 and S3, the nonvanishing $\gamma_0$ component induced around the core breaks axial symmetry but satisfies $\gamma_0(\rho,\theta) = \gamma^{\ast}_0(\rho,-\theta)$. In addition, no $\gamma_{\pm 1}$ components are induced. Hence, the HQV maintains the $P_{3}$ symmetry. Table~\ref{table1} summarizes the possible classes of vortices in terms of these discrete symmetries. In the previous work~\cite{Masaki:2019rsz}, we demonstrated that the $o$ vortex and $\varv$ vortex are the only stable solution of the self-consistent equations, while the no stable solutions of the $u$-, $w$, and $u \varv w$ vortex are obtained in the $D_{4}$-BN state with the constraints of the axial symmetry.

The mirror reflection with respect to the $xy$-plane is defined as the following transformation of the real-space, the momentum, and the spin variables: $(x,y,z)\rightarrow (x,y,-z)$, $(k_{x},k_{y},k_{z})\rightarrow (k_{x},k_{y},-k_{z})$, and $(\sigma_{x},\sigma_{y},\sigma_{z})\rightarrow (-\sigma_{x},-\sigma_{y},\sigma_{z})$. For systems with spatial uniformity along the $z$ axis, only the momentum and spin variables transform under the mirror reflection since the BdG Hamiltonian does not depend on $z$. The mirror reflection operator in the spin space, $M_{xy}$, is simply given by $\hat{M}_{xy}=i\hat{\sigma}_{z}$. 
When the $^{3}P_{2}$ superfluid order parameter is invariant under the mirror reflection symmetry, the operator acts as
\begin{align}
\hat{M}_{xy} \hat{\Delta}({\bm k},\theta) \hat{M}_{xy}^{\mathrm{tr}} = \eta\hat{\Delta}(k_{x},k_{y},-k_{z},\theta).
\end{align}
As the parity under the mirror reflection, $\eta=\pm$, is compensated by the discrete $\mathrm{U}(1)$ rotation, the mirror operator in the Nambu space is given as
$\check{\mathcal{M}}^{\eta}_{xy} = \mathrm{diag}(\hat{M}_{xy},\eta \hat{M}^{\ast}_{xy})$.
The BdG Hamiltonian \eqref{eq:bdg} is invariant under mirror reflection symmetry, \ie, 
\begin{align}
\check{\mathcal{M}}^{\eta}_{xy}\check{\mathcal{H}}({\bm k},\theta) \check{\mathcal{M}}^{\eta \dag}_{xy} = \check{\mathcal{H}}(k_{x},k_{y},-k_{z},\theta).
\label{eq:mirrorBdG}
\end{align}
The mirror reflection operator thus acts on the symmetric traceless tensor $\mathcal{A}_{\mu i}$ as 
\begin{align}
M_{xy}: ~ \mathcal{A}_{\mu i} \rightarrow \begin{pmatrix}
-\mathcal{A}_{xx} & -\mathcal{A}_{xy} & \mathcal{A}_{xz} \\
-\mathcal{A}_{xy} & -\mathcal{A}_{yy} & \mathcal{A}_{yz} \\
\mathcal{A}_{xz} & \mathcal{A}_{yz} & -\mathcal{A}_{zz}
\end{pmatrix}.
\label{eq:mirror}
\end{align}
Therefore, the BdG Hamiltonian is invariant under the mirror reflection if $\mathcal{A}_{xz} = \mathcal{A}_{yz}=0$, \ie, $\gamma_{M=\pm 1} = 0$. This condition can be satisfied by the $o$ vortex and the non-Abelian HQVs. In Table~\ref{table1}, we summarize the symmetry and topology of the several classes of topological line defects in the $D_{4}$-BN states.

%=================================
\subsection{S5.2. Topological invariants for vortex-bound states}

One of the topological invariants appropriate for line defects in the Altland--Zirnbauer symmetry class D is the second Chern number~\cite{teo,tsutsumiPRB15}. Consider the BdG Hamiltonian at an infinite point from a vortex, $\check{\mathcal{H}}({\bm k},\theta)$, which is obtained from Eq.~\eqref{eq:bdg} on the four-dimensional space $({\bm k},\theta)$ as $\check{\mathcal{H}}({\bm k},\theta)=\lim _{\rho \rightarrow \infty}\check{\mathcal{H}}({\bm k},{\bm R})$. The second Chern number is given by 
\begin{align}
\mathrm{Ch}_{2} = \frac{1}{8\pi^{2}} \int _{S^{3}\times S^{1}} \mathrm{tr}\left[
\mathcal{F}\wedge \mathcal{F}
\right],
\end{align}
where $\mathcal{F} = d{A} + {A}\wedge {A}$ is the non-Abelian field strength in the four-dimensional base space. Let $\ket{u_n({\bm k},\theta)}$ be the $n$th occupied eigenstate of $\check{\mathcal{H}}({\bm k},\theta)$. The non-Abelian Berry's connection, $A_{nm}^{\alpha}$, is defined with the eigenstates as 
\begin{align}
A^{\alpha}_{nm} = -i\braket{u_{n}({\bm k},\theta)| \partial _{\alpha} u_{m}({\bm k},\theta)}.
\end{align}
Here we introduce $(\partial_{1},\partial_{2},\partial_{3},\partial_{4}) \equiv (\partial_{k_{x}},\partial_{k_{y}},\partial_{k_{z}}, \partial_{\theta})$ with $\alpha = 1, 2,3,4$. Owing to the particle-hole symmetry, the value of the second Chern number in the class D is even, \ie, $\mathrm{Ch}_{2}=2\mathbb{Z}$~\cite{teo}. Hence, the nontrivial value of $\mathrm{Ch}_{2}$ ensures the appearance of paired Majorana fermions at a topological line defect. It has also been pointed out that the second Chern number vanishes in the presence of the $P_{1}$ symmetry~\cite{tsutsumiPRB15}, and thus the $o$ vortex has $\mathrm{Ch}_{2}=0$. We also find that the Chern number in the other vortices is trivial, $\mathrm{Ch}_{2}=0$.

When the BdG Hamiltonian \eqref{eq:bdg2} satisfies Eq.~\eqref{eq:mirrorBdG} and keeps the mirror reflection symmetry, one can introduce a topological number called the mirror
$\mathbb{Z}_{2}$ number: Since the mirror invariant BdG Hamiltonian commutes with the mirror operator, it can be block diagonal by using the eigenvalues of the mirror operator. Here we should emphasize that the zero-energy vortex-bound state can be the Majorana zero mode only for $\check{\mathcal{M}}_{xy}^{\eta=-}$ while the mirror $\mathbb{Z}_{2}$ number can be defined for both of the two possible mirror symmetries $\check{\mathcal{M}}_{xy}^{\eta = \pm}$. 
This is because the particle-hole symmetry is still supported within each subsector for $\check{\mathcal{M}}_{xy}^{-}$ but not for $\check{\mathcal{M}}_{xy}^{+}$. 
In terms of the Altland--Zirnbauer symmetry classes, 
each subsector for $\check{\mathcal{M}}_{xy}^{-}$ belongs to the class D as well as spinless chiral superconductors.
In contrast, for $\check{\mathcal{M}}_{xy}^{+}$, each mirror subsector does not maintain the particle-hole symmetry, and its Hamiltonian belongs to the class A, as well as a quantum Hall state. Thus, the zero-energy states behave as Majorana fermions only for $\check{\mathcal{M}}^{-}_{xy}$ and only Dirac fermions can be obtained for $\check{\mathcal{M}}^{+}_{xy}$. For $\check{\mathcal{M}}^{-}_{xy}$, if the $\mathbb{Z}_{2}$ invariant is odd, the vortex can host a single Majorana zero mode which behaves as a non-Abelian (Ising) anyon~\cite{ueno,sato14,tsutsumiJPSJ13}.

As mentioned in Eq.~\eqref{eq:mirror}, the BdG Hamiltonian for the $o$ vortex and the non-Abelian HQVs is invariant under $\check{\mathcal{M}}_{xy}^{-}$ if $\mathcal{A}_{xz} = \mathcal{A}_{yz}=0$, \ie, $\gamma_{M=\pm 1} = 0$. On the mirror reflection invariant plane at ${\bm k}_{\mathrm{M}}\equiv (k_{x},k_{y},k_{z}=0)$, the BdG Hamiltonian is commutable with $\check{\mathcal{M}}_{xy}^{-}$, $[\check{\mathcal{H}}({\bm k}_{\mathrm{M}},\theta),\check{\mathcal{M}}_{xy}^{-}]=0$. Hence, the BdG Hamiltonian is block-diagonal for the $2\times 2$ submatrices, $\tilde{\mathcal{H}}_{\lambda}({\bm k}_{\mathrm{M}},\theta)$, 
in terms of the eigenvalues of the mirror operator $\lambda = \pm i$
as 
\begin{align}
\check{\mathcal{H}}({\bm k}_{\mathrm{M}},\theta) = \bigoplus_{\lambda}\tilde{\mathcal{H}}_{\lambda}({\bm k}_{\mathrm{M}},\theta)
\end{align}
where $\tilde{\mathcal{H}}_{\lambda}({\bm k}_{\mathrm{M}},\theta)$ is still subject to the particle-hole symmetry with $\tilde{\mathcal{C}}=\tilde{\tau}_{x}K$ ($\tilde{\tau}_{\mu=x,y,z}$ are the $2\times2$ Pauli matrix in a reduced subspace) and belongs to the class D in each mirror subsector. Then, the BdG Hamiltonian, $\tilde{\mathcal{H}}_{\lambda}$, is obtained on the base space $({\bm k}_{\mathrm{M}},\theta) \in S^{2}\times S^{1}$, which is composed of the two-dimensional mirror-invariant momentum space and a loop enclosing the line defect on the mirror invariant plane in the real space. The topological invariant for the Altland--Zirnbauer symmetry class D is the $\mathbb{Z}_{2}$ number~\cite{teo,qiPRB08}. The $\mathbb{Z}_{2}$ number is obtained by the dimensional reduction of the second Chern number in the suspension of the base space, $\Sigma(S^{2}\times S^{1})$, which is given by the integral of the Berry curvature $\mathcal{F}\wedge\mathcal{F}$ over the suspension. The $\mathbb{Z}_{2}$ number in each mirror subsector is 
\begin{align}
\nu_{\lambda} = \left(\frac{i}{\pi} \right)^{2} \int _{S^{2}\times S^{1}} Q_{3}~ \mod 2,
\end{align}
where $Q_{3}=\mathrm{tr}[\tilde{A}d\tilde{A} + \frac{2}{3}\tilde{A}^{3}]$ is the Chern-Simons form and the Berry connection $\tilde{A}$ is obtained from the eigenstates on the mirror reflection invariant plane.
For the vortex preserving the mirror symmetry, the $\mathbb{Z}_{2}$ invariant in the $\lambda$ subsector is 
\begin{align}
\nu_{\lambda} = \ell \kappa_{\lambda}~ \mod 2,
\end{align}
where $\kappa_{\lambda}$ is the vorticity of the order parameter in the $\lambda$ subsector, and $\ell$ is the first Chern number characterizing the bulk topology of the mirror invariant plane and $\ell=\pm 1$ for $\lambda = \pm i$. On the basis of the total angular momentum $\Gamma_{M}$,  the vorticity $\kappa_{\lambda}$ is equivalent to the vorticity $\kappa_{M}$ of $J_{z}=M$ Cooper pairs. Hence, the mirror-protected $\mathbb{Z}_{2}$ invariants for an isolated HQV with $(\kappa,n)=(1/2,1/4)$ and $(1/2,-1/4)$ are $(\nu_{\lambda=+i},\nu_{\lambda=-i})=(0,1)$ and $(1,0)$, respectively. Hence, the zero energy state bound at each HQV is protected by the $\mathbb{Z}_{2}$ invariant regardeless of the intervortex distance $d_{\mathrm{V}}$. In Ref.~\cite{Masaki:2019rsz}, we found that a pair of the zero energy states exist in the $o$ vortex. The topological protection of a pair of zero energy states are also characterized by the $\mathbb{Z}_{2}$ invariants as $(\nu_{\lambda=+i},\nu_{\lambda=-i})=(1,-1)$.

Finally, we introduce another topological invariant associated with the chiral symmetry~\cite{tsutsumiJPSJ13,tsutsumiPRB15,Masaki:2019rsz,mizushimaJPSJ16}. Consider the vortex preserving the $P_{3}$ symmetry, that is, the magnetic $\pi$ rotation symmetry. Then, one can construct the chiral operator $\check{\Gamma}$ as a combination of $\mathcal{C}$ in Eq.~\eqref{eq:phs} and $P_{3}$ operator, and the BdG Hamiltonian $\check{\mathcal{H}}({\bm k},\theta)$ preserves the chiral symmetry, 
\begin{align}
\{\check{\Gamma},\check{\mathcal{H}}({\bm k},\theta) \}=0, ~ \hat{\Gamma}^{2} = +1
\end{align}
for $k_{y}=k_{z}=0$. As long as the chiral symmetry is preserved, one can define the one-dimensional winding number for $\theta$ as 
\begin{align}
w_{\mathrm{1d}}(\theta) = -\frac{1}{4\pi i} \int dk_{x} \mathrm{tr}\left[
\check{\Gamma}\check{\mathcal{H}}({\bm k},\theta)\partial _{k_{x}} \check{\mathcal{H}}({\bm k},\theta)
\right]_{k_{y}=k_{z}=0}.
\end{align}
The winding number for the $o$-vortex state is $w_{\mathrm{1d}}(\theta=0) = 2$ and $w_{\mathrm{1d}}(\theta=\pi) = -2$, regardless of the boundary condition. Then, the topological invariant is defined as the difference of $w_{\mathrm{1d}}(\theta)$, $w_{\mathrm{1d}} = \frac{w_{\mathrm{1d}}(0)-w_{\mathrm{1d}}(\pi)}{2} = 2$. This ensures the presence of the two zero energy states at $k_{z}=0$. Similarly, the winding number describes the topological protection of the zero energy state in an isolated HQV. We thus conclude that the zero energy state and its non-Abelian anyonic nature in the HQVs are protected by both the mirror $\mathbb{Z}_{2}$ invariants $\nu_{\lambda}$ and the winding number $w_{\mathrm{1d}}$.

In summary, a single Majorana zero mode bound at the HQV in the $D_4$-BN state is protected by the two topological invariants associated with the mirror reflection symmetry and chiral symmetry. These topological invariants ensure that the Majorana zero mode behaves as a non-Abelian (Ising) anyon. Therefore, the HQV hosts two-fold non-Abelian anyons characterized by both Majorana fermions and a non-Abelian first homotopy group. Both the symmetries can be preserved and topological invariants are unchanged even if two HQVs form a bound molecule and the magnetic field is applied along the vortex line. Therefore, the Majorana zero mode and its non-Abelian anyonic nature are tolerant against a magnetic field. 

\bibliographystyle{apsrev4-2}
\bibliography{reference}